\documentclass[10pt,twocolumn,letterpaper]{article}

\usepackage{cvpr}
\usepackage{times}
\usepackage{epsfig}
\usepackage{graphicx}
\usepackage{amsmath}
\usepackage{amssymb}

\usepackage{epsfig}
\usepackage{subfig}
\usepackage{subfig}
\usepackage{pifont}
\usepackage{multirow}
\usepackage{color}

\usepackage[breaklinks=true,bookmarks=false]{hyperref}

\cvprfinalcopy 

\def\cvprPaperID{84} 

\ifcvprfinal\fi
\begin{document}

\title{Adaptive Weighted Attention Network with Camera Spectral Sensitivity Prior for Spectral Reconstruction from RGB Images}

\author{
Jiaojiao Li$^{1}\thanks{These authors contribute equally to this work.}$ \hspace*{0.5cm} Chaoxiong Wu$^{1}\footnotemark[1]$ \hspace*{0.5cm} Rui Song$^1$ \hspace*{0.5cm} Yunsong Li$^1$ \hspace*{0.5cm} Fei Liu$^2$
\\$^1$The State Key Laboratory of Integrated Service Networks, Xidian University, Xi’an 710071, China
\\$^2$School of Physics and Optoelectronic Engineering, Xidian University, Xi’an 710071, China
\\{\tt\small \{jjli, rsong, ysli, feiliu\}@xidian.edu.cn, cxwu@stu.xidian.edu.cn}
}

\maketitle
\thispagestyle{empty}

\begin{abstract}
Recent promising effort for spectral reconstruction (SR) focuses on learning a complicated mapping through using a deeper and wider convolutional neural networks (CNNs). Nevertheless, most CNN-based SR algorithms neglect to explore the camera spectral sensitivity (CSS) prior and interdependencies among intermediate features, thus limiting the representation ability of the network and performance of SR. To conquer these issues, we propose a novel adaptive weighted attention network (AWAN) for SR, whose backbone is stacked with multiple dual residual attention blocks (DRAB) decorating with long and short skip connections to form the dual residual learning. Concretely, we investigate an adaptive weighted channel attention (AWCA) module to reallocate channel-wise feature responses via integrating correlations between channels. Furthermore, a patch-level second-order non-local (PSNL) module is developed to capture long-range spatial contextual information by second-order non-local operations for more powerful feature representations. Based on the fact that the recovered RGB images can be projected by the reconstructed hyperspectral image (HSI) and the given CSS function, we incorporate the discrepancies of the RGB images and HSIs as a finer constraint for more accurate reconstruction. Experimental results demonstrate the effectiveness of our proposed AWAN network in terms of quantitative comparison and perceptual quality over other state-of-the-art SR methods. In the NTIRE 2020 Spectral Reconstruction Challenge, our entries obtain the 1st ranking on the ``Clean" track and the 3rd place on the ``Real World" track. Codes are available at {\color{magenta} https://github.com/Deep-imagelab/AWAN}.
\end{abstract}

\begin{figure*}[!tbp]
	\vskip 0.05in
	\centering
	\scalebox{1}
	{
		\begin{tabular}{@{}c@{}c@{}c@{}c@{}c@{}c@{}c@{}c}
			&${\text{Arad~\cite{arad2016sparse}}}$ & $\text{Galliani~\cite{galliani2017learned}}$ &$\text{Yan~\cite{yan2018accurate}}$ &$\text{Stiebel~\cite{stiebel2018reconstructing}}$&$\text{HSCNN-R~\cite{shi2018hscnn+}}$&$\text{Ours}$ \\
			&{\includegraphics[width=2.41cm,height=2.41cm]{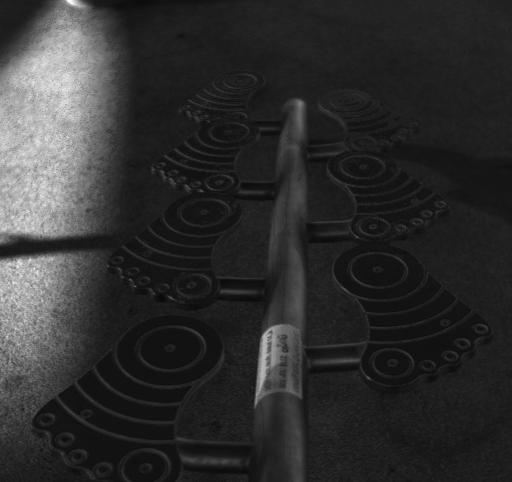}} \
			&{\includegraphics[width=2.41cm,height=2.41cm]{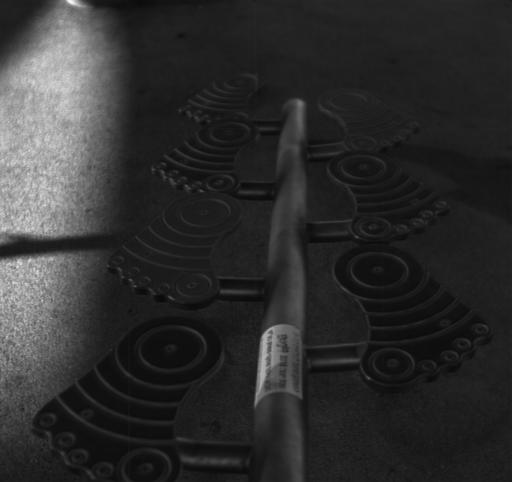}} \
			&{\includegraphics[width=2.41cm,height=2.41cm]{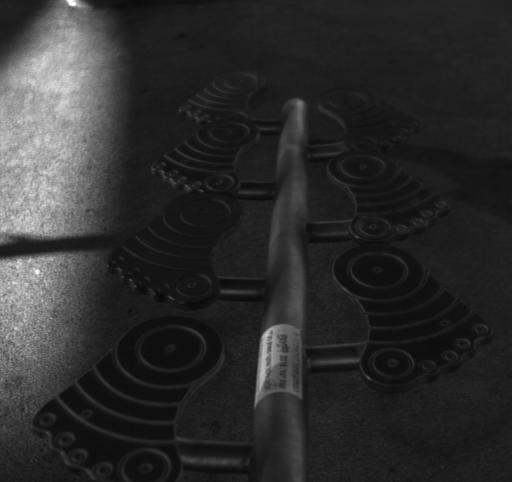}} \
			&{\includegraphics[width=2.41cm,height=2.41cm]{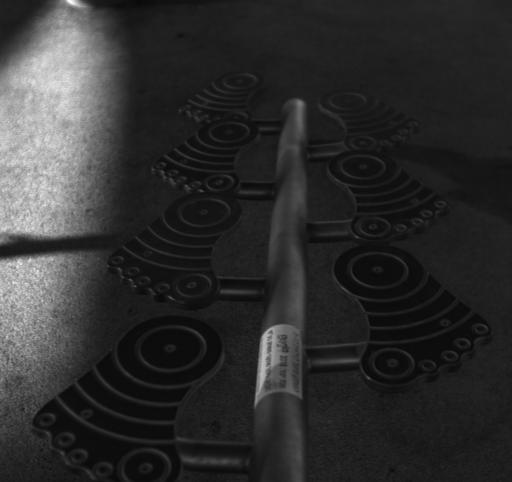}} \
			&{\includegraphics[width=2.41cm,height=2.41cm]{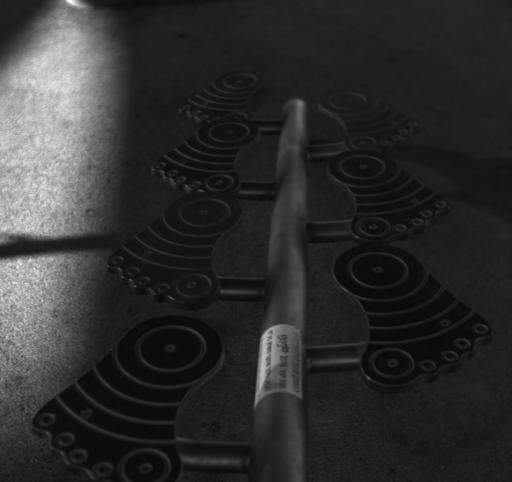}} \
			&{\includegraphics[width=2.41cm,height=2.41cm]{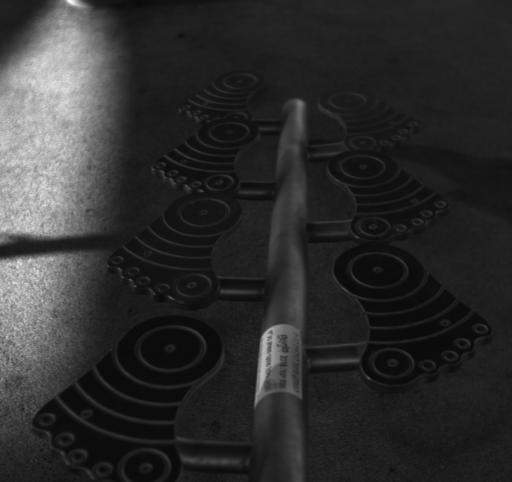}} \\
			&{\includegraphics[width=2.41cm,height=2.41cm]{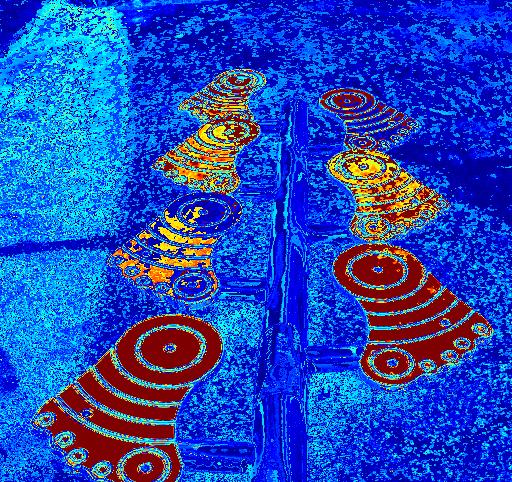}} \
			&{\includegraphics[width=2.41cm,height=2.41cm]{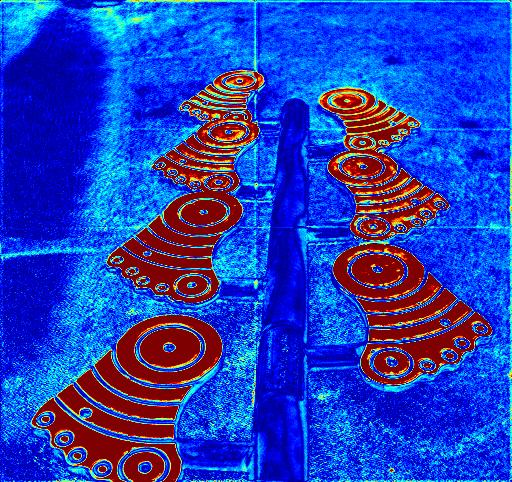}} \
			&{\includegraphics[width=2.41cm,height=2.41cm]{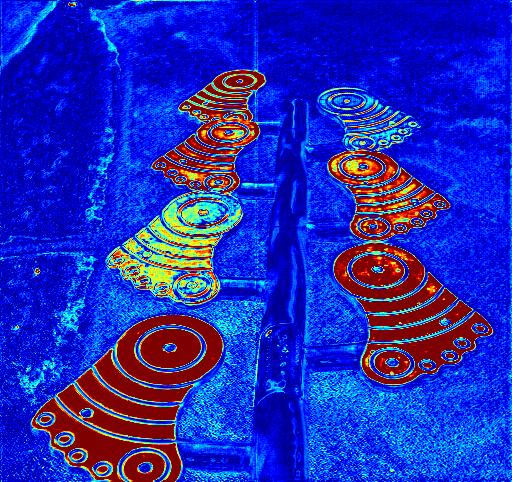}} \
			&{\includegraphics[width=2.41cm,height=2.41cm]{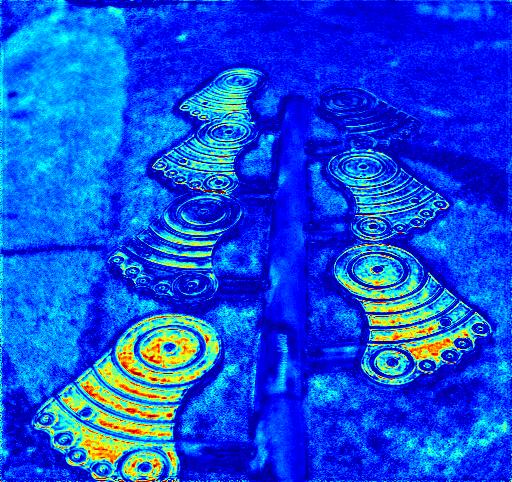}} \
			&{\includegraphics[width=2.41cm,height=2.41cm]{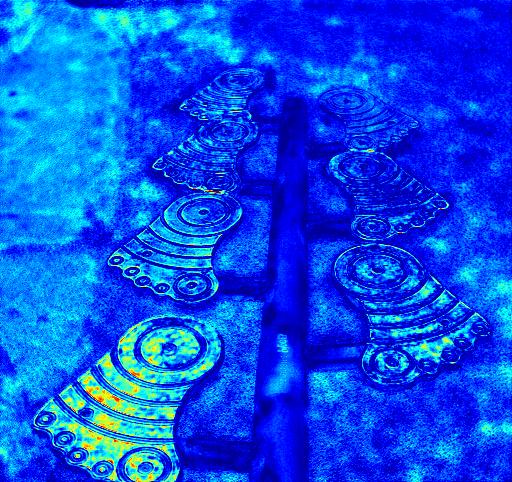}} \
			&{\includegraphics[width=2.41cm,height=2.41cm]{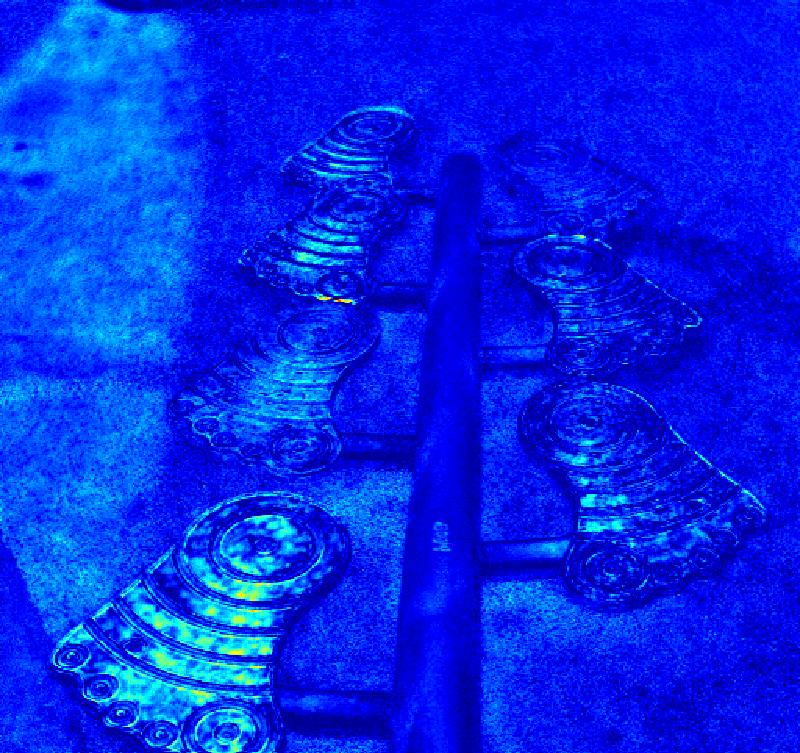}}\
			&{\includegraphics[width=0.2cm,height=2.41cm]{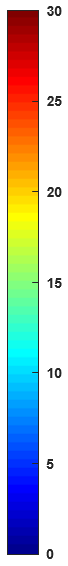}} 
			\\
		\end{tabular}
	}
	\caption{The Visual results of the 18-th band and the reconstruction error images of an HSI chosen from validation set of NTIRE2020 ``Clean'' track. The error images are the heat maps of mean relative absolute error (MRAE) between the ground truth and the recovered HSI. our approach obtains more precise HSI and better recovered quality over other SR methods.}
	\label{figure1}	
	\vskip -0.14in
\end{figure*}

\section{Introduction}
\label{Introduction}
Hyperspectral imaging records the reflectance or transmittance of objects and the acquired hyperspectral images (HSIs) typically have a multitude of spectral bands ranging from the infrared spectrum to ultraviolet spectrum. The rich spectral signatures have been widely explored to various tasks \eg, face recognition, image classification and anomaly detection\cite{pan2003face,li2018hyperspectral,stein2002anomaly}. However, capturing such HSIs containing plentiful spectral information with high spatial/temporal resolution is time consuming due to the limitations of the imaging technology, hence ineluctably preventing the application scope of HSIs.

One way to solve this problem is to develop scan-free or snapshot hyperspectral devices based on compressed sensing and computational reconstruction, for instance, computed tomography imaging spectrometers (CTIS) \cite{descour1995computed}, hybrid RGB-HS systems \cite{kwon2015rgb} and aperture masks \cite{cao2011prism} etc. Nevertheless, these acquisition systems still rely on expensive hardware devices. Another effective way is to generate such HSIs through recovering the lost spectral information from a given RGB image, defined as spectral reconstruction (SR) or spectral super-resolution. However, this inverse process is severely ill-posed since amounts of HSIs can project to any RGB input. To make the problem resolvable, a large number of SR approaches have been proposed, roughly divided into two categories: early sparse/shallow learning methods \cite{arad2016sparse} and recent deep CNN-based models \cite{galliani2017learned,arad2018ntire}.

The early researchers mainly concentrate on building sparse coding or relatively shallow learning models from a specific hyperspectral prior to fulfill spectral super-resolution \cite{robles2015single,arad2016sparse,jia2017rgb,aeschbacher2017defense}. Nonetheless, these methods are restricted to perform well on images in specific domains owing to the poor expression capacity and the limited generalizability. In recent years, as CNNs have achieved remarkable success in many computer vision tasks, a series of CNN-based SR models are also presented to learn a mapping function from a single RGB image to its corresponding HSI \cite{kaya2019towards,yan2018accurate, arad2018ntire,shi2018hscnn+,stiebel2018reconstructing,xiong2017hscnn}. Besides, the self-attention mechanism for capturing the long range dependencies is adopted for SR \cite{Miao_2019_ICCV}. Although promising performances have been implemented in SR, the existing approaches based on CNNs still involve with some disadvantages. Most of CNN-based SR methods devote to design deeper or wider network architectures to acquire a more advanced feature expression, lacking the exploration of rich contextual information and interdependencies among intermediate features, therefore restricting the discriminative learning of CNNs. Additionally, the existing CNN-based SR models invariably accomplish a complicated RGB-to-HSI mapping function and rarely consider to integrate camera spectral sensitivity (CSS) prior into SR for more accurate reconstruction.

To address these issues, a novel deep adaptive weighted attention network (AWAN) for SR is proposed in this paper. In specific, the backbone architecture of our AWAN network is constituted of multiple dual residual attention blocks (DRAB), in which the long and short skip connections form the dual residual learning to allow abundant low-frequency information to be bypassed to enhance feature correlation learning. Moreover, we present a trainable adaptive weighted channel attention (AWCA) module for better modeling channel-wise dependencies. Our AWCA module adaptively reallocates channel-wise feature responses by exploiting adaptive weighted feature statistics instead of average-pooled ones. Besides, for more powerful feature representation, a patch-level second-order non-local (PSNL) module is developed to capture long-range spatial contextual information by the second-order non-local operations. Based on the fact that the recovered RGB images can be generated through employing the known CSS function to reconstructed HSI, we incorporate the discrepancies of the RGB images and HSIs as a finer constraint for more accurate reconstruction. As shown in Fig. \ref{figure1}, our approach obtains more precise HSI and better reconstruction quality over other different SR methods.

The main contributions of this paper are summarized as follows:	
\begin{enumerate}
	\item A novel deep adaptive weighted attention network (AWAN) for SR is presented. Experimental results  demonstrate the effectiveness of the proposed AWAN in terms of quantitative comparison and perceptual quality. In the NTIRE 2020 Spectral Reconstruction Challenge \cite{Arad_2020_CVPR_Workshops}, our entries obtain the 1st ranking on the ``Clean" track and the 3rd place only 1.59106e-4 more than the 1st on the ``Real World" track.
	\item We propose an adaptive weighted channel attention (AWCA) module to adaptively recalibrate channel-wise feature responses by exploiting the adaptive weighted feature statistics instead of average-pooled ones. Such AWCA module allows our network to selectively emphasize informative features and boost discriminant learning power.
	\item We develop a patch-level second-order non-local (PSNL) module to capture long-range spatial contextual information via second-order non-local operations for more powerful feature representations.  
	\item The CSS function prior is integrated into the SR Loss process for more accurate reconstruction through incorporating the discrepancies of the RGB images and HSIs as a finer constraint.
\end{enumerate}

\begin{figure*}
	\centering
	\includegraphics[width=0.85\textwidth]{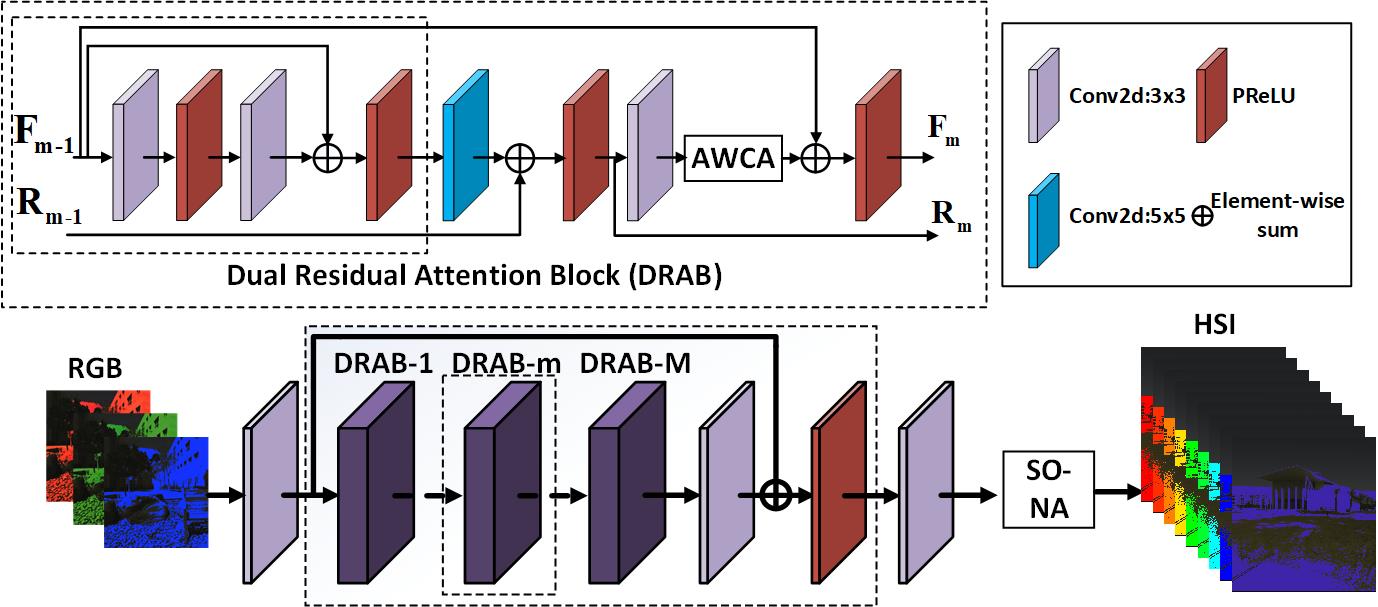}
	\caption{Network architecture of our adaptive weighted attention network (AWAN). ${\textbf{F}_\textbf{{m-1}}}$ and ${\textbf{F}_\textbf{m}}$ denote direct input anf output of the ${m}$-th DRAB. ${\textbf{R}_\textbf{{m-1}}}$ and ${\textbf{R}_\textbf{m}}$ denote residual input anf output of the ${m}$-th DRAB.} 
	\label{figure2}
	\vskip -0.14in
\end{figure*}

\section{Related Work}
\label{Related Work}

In the past few years, an increasing amount of algorithms for SR have been proposed, including specific acquisition systems~\cite{kwon2015rgb,cao2011prism}, sparse/shallow learning methods~\cite{robles2015single,arad2016sparse,jia2017rgb,aeschbacher2017defense} and CNN-based models~\cite{galliani2017learned,yan2018accurate,fu2018joint,nie2018deeply,arad2018ntire,shi2018hscnn+,zhang2019pixel,stiebel2018reconstructing,can2018efficient,alvarez2017adversarial,koundinya20182d}. Here we summarize some CNN-based SR works and the attention mechanism without enumerating them all due to space limitation.

\textbf{CNN-based SR models.} Recently, CNN-based SR methods have been widely studied and developed with the great success of CNNs in computer vision tasks. Typically, these methods formulate SR as an image-to-image regression problem and learn a deep mapping function from three-dimensional RGB pixel values to high-dimensional hyperspectral signals. Initially, Galliani \etal~\cite{galliani2017learned} and Xiong \etal~\cite{xiong2017hscnn} trained an end-to-end CNN for SR, which achieved unprecedented results. Later, Arad \etal~\cite{arad2018ntire} organized the NTIRE 2018 Spectral Reconstruction Challenge and a plent of excellent algorithms were proposed. For instance, Shi \etal~\cite{shi2018hscnn+} proposed a deep residual network HSCNN-R consisting of adapted residual blocks. To further improve performance, they designed a deeper HSCNN-D model based on a densely-connected structure with a novel fusion scheme. Stiebel \etal~ \cite{stiebel2018reconstructing} introduced modified U-net from the semantic segmentation to this task and won the 4th place in the spectral reconstruction competition. To increase the flexibility of the network for learning the pixel-wise mapping, Zhang \etal~\cite{zhang2019pixel} completed the RGB-to-HSI mapping using a pixel-aware deep function-mixture network composing of a couple of function-mixture blocks.

\textbf{Attention mechanism.} In general, attention mechanism can be viewed as a tool to redistribute available information and focus on the salient components of an image \cite{vaswani2017attention}, which has already played an important role in the current computer vision society, such as video classification, super-resolution and scene segmentation \cite{wang2018non,dai2019second,fu2019dual} etc. In specific, Xia \etal~\cite{xia2019second} presented a novel attention module in spatial domain incorporating non-local operations with second-order statistics in CNN to extract contextual dependencies directly for person re-identification and gain superior performance. Due to that the realization of non-local operations in whole image is time consuming, we develop a patch-level second-order non-local (PSNL) module to reduce the computational burden. Hu \etal~\cite{hu2018squeeze} proposed a squeeze-and-excitation (SE) block in channel domain to model channel-wise feature correlations for image classification. However, this attention module adaptively reallocates channel-wise feature responses by exploiting global average-pooled statistics, indicating that it treats equally across spatial locations without consideration of different importance degree of them, resulting to hinder the representational power of CNNs. Therefore, we propose a novel deep adaptive weighted attention network (AWAN) by exploring adaptive weighted feature statistics for stronger feature representations.

\begin{figure*}
	\centering
	\includegraphics[width=0.85\textwidth]{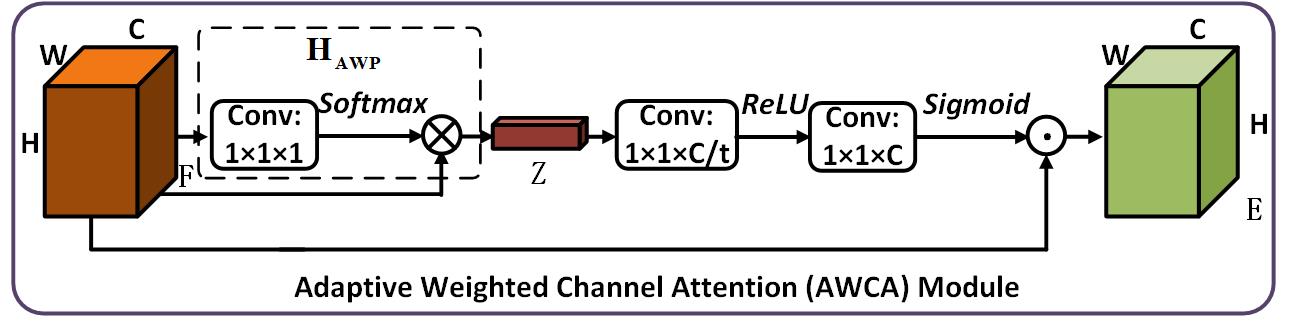}
	\caption{Diagram of adaptive weighted channel attention (AWCA) module. $\odot$ denotes element-wise multiplication.} 
	\label{figure3}
	\vskip -0.14in
\end{figure*}

\section{Our Proposed Method}
\label{Our Proposed Method}

\subsection{Network Architecture}
\label{Network Architecture}
The overall architecture of proposed AWAN is illustrated in Fig. \ref{figure2}. Firstly, we employ an individual convolutional layer to extract the shallow features from the RGB input. Then we stack ${M}$ dual residual attention blocks (DRABs) to form a deep network for the deep feature extraction. To eliminate the problem of gradient vanishing and explosion in the very deep network, the global residual connection is adopted. Each DRAB consists of a fundamental residual module \cite{he2016deep} and additional paired convolutional operations with a large (${5 \times 5}$) and small size (${3 \times 3}$) kernels, in which the long and short skip connections form the dual residual learning in the block. This type of residual in residual structure makes the best of exploiting the potential of pairwise operations by increasing the interaction between the basic residual blocks. Also, such module can allow abundant low-frequency information of the original RGB images to be bypassed and utilized adequately, which enhances the feature correlation learning greatly. Different from the work \cite{liu2019dual}, batch normalization is not applied in our paper, since the normalization limits the strength of the network to learn correlations between the spectral distribution and the local spatial intensities for SR task, which can further reduce its robustness to variations in the intensity range of an HSI. Besides, we choose Parametric Rectified Linear Unit (PReLU) rather than ReLU as activation function to introduce more nonlinear and accelerate convergence.

\subsection{Adaptive Weighted Channel Attention (AWCA)}
\label{Adaptive Weighted Channel Attention (AWCA)}
Extracting interdependencies among intermediate features is indispensable for strengthening discriminant learning power of CNNs. SE block \cite{hu2018squeeze} is proposed to adaptively recalibrate channel-wise feature responses by explicitly modelling interdependencies between channels. However, it treats equally across spatial locations by exploiting global average-pooled statistics in the squeeze process, thus preventing the representational capacity of CNNs. For more powerful feature correlation learning, an adaptive weighted channel attention (AWCA) module is proposed to selectively emphasize the informative features by exploring adaptive weighted feature statistics.

Given an intermediate feature map group denoted as $\textbf{F}=[\textbf{f}_1, \textbf{f}_2, \cdots ,\textbf{f}_c, \cdots , \textbf{f}_C]$ containing $C$ feature maps with size of $H \times W$ and then reshape \textbf{F} to ${R^{C \times (H\times W)}}$. We exploit one convolutional layer to learn the adaptive weighted matrix $\textbf{Y}\in R^{1\times H\times W}$ and reshape \textbf{Y} to ${R^{(H\times W)\times1}}$. Then we apply a softmax layer to normalize \textbf{Y} and multiply \textbf{F} with \textbf{Y}. As shown in Fig. \ref{figure3}, we define the above process as adaptive weighted pooling $H_{AWP}(\cdot)$
\begin{equation}
\label{equa2}
\textbf{Z}=H_{AWP}(\textbf{F})
\end{equation}
where $\textbf{Z}=[\textbf{z}_1, \textbf{z}_2, \cdots , \textbf{z}_C](\textbf{Z}\in R^{C\times1})$ is channel-wise descriptors. To make use of the aggregated information \textbf{Z} by adaptive weighted pooling, we adopt a simple gating mechanism with sigmoid function, where the output dimension of the first convolutional layer is $R^{(C/t)\times1\times1}$ and the output size of the second convolutional layer is $R^{C\times1\times1}$. $t$ is the reduction ratio. The final channel map is computed as:
\begin{equation}
\label{equa3}
V = \delta (W_2(\sigma (W_1(\textbf{Z})))) 
\end{equation}
where $W_1$ and $W_2$ are the weight set of two convolutional layers. $\delta(\cdot)$ and $\sigma(\cdot)$ denote the sigmoid and ReLU activation functions. Then we assign channel attention map $V=[v_1,v_2,\cdots,v_c,\cdots,v_C]$ to rescale the input $\textbf{F}$
\begin{equation}
\label{equa4}
\textbf{e}_c = v_c \cdot \textbf{f}_c
\end{equation}
where $v_c$ and $\textbf{f}_c$ are the scaling factor and feature map of the $c$-th channel. $\textbf{E}=[\textbf{e}_1, \textbf{e}_2, \cdots, \textbf{e}_c, \cdots , \textbf{e}_C]$
is output feature map of AWCA module. Embedded with AWCA block, the proposed DRAB module can adjust channel-wise feature recalibration adaptively to boost representational learning of the network.

\begin{figure*}
	\centering
	\includegraphics[width=0.85\textwidth]{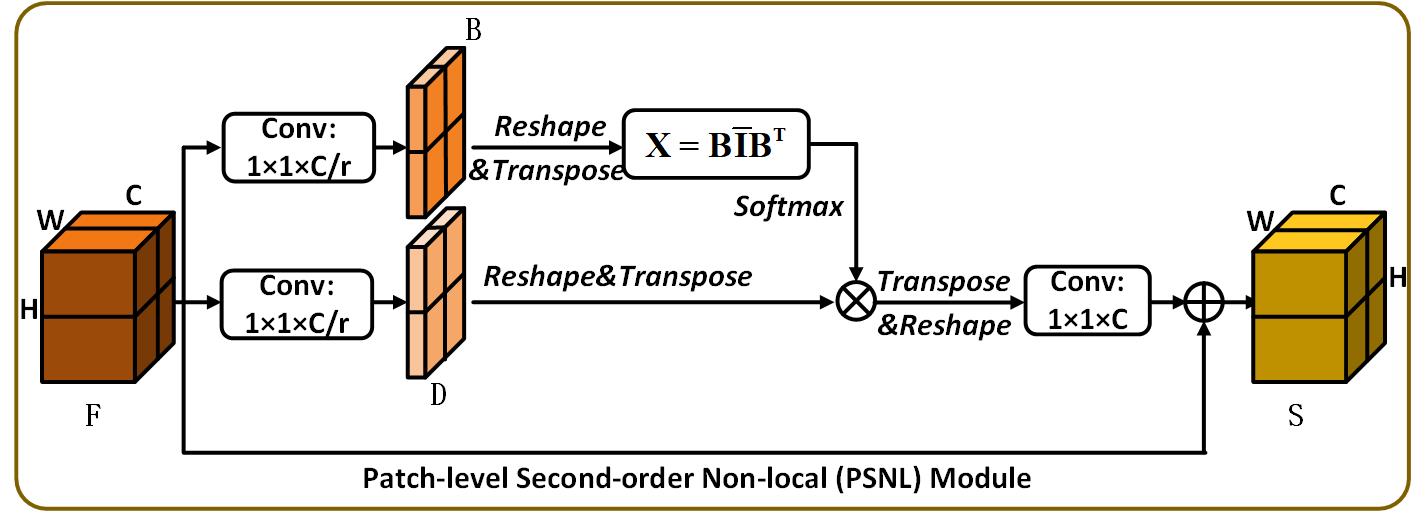}
	\caption{Diagram of patch-level second-order non-local (PSNL) module. $\otimes$ denotes matrix multiplication.}
	\label{figure4}
	\vskip -0.14in
\end{figure*}

\subsection{Patch-level Second-order Non-local (PSNL)}
\label{Patch-level Second-order Non-local (PSNL)}
The non-local neural block \cite{wang2018non} was proposed to capture the long-range dependencies throughout the entire image. Meanwhile, recent works \cite{dai2019second,xia2019second} have indicated that second-order statistics is an effective tool for more discriminative representations of CNNs. However, non-local operations need huge computational burden. To decrease computational cost and model distant region relationships simultaneously, we develop a patch-level second-order non-local (PSNL) module. Fig.\ref{figure4} depicts an illustration of PSNL module. Given a feature map $\textbf{F}\in R^{C\times H\times W}$, we split it into four sub-feature maps $\textbf{F}_k\in R^{C\times h\times w} (k=1,2,3,4;h=H/2;w=W/2)$ along spatial dimension, each of which is processed by the subsequent PSNL module.

Firstly, we feed the feature map $\textbf{F}_k\in R^{C\times h\times w}$ into a $1\times1$ convolutional layer with output channel $=C/r$ to produce two new feature maps $\textbf{B}_k$ and $\textbf{D}_k$, respectively. Then we reshape and transpose them to $R^{(h\times w)\times C/r}$. The covariance matrix can be computed using \textbf{B} as
\begin{equation}
\label{equa5}
\textbf{X}_k=\textbf{B}_k \overline{\textbf{I}} \textbf{B}_k^{\textbf{T}}
\end{equation}
where $\overline{\textbf{I}}=\frac{1}{n}\left(\textbf{I}-\frac{1}{n} \textbf{1}\right)$, $n=h\times w$ and $\textbf{X}_k\in R^{n\times n}$. \textbf{I} and \textbf{1} represent the $n\times n$ identity matrix and matrix of all ones. $\textbf{X}_k$ is the spatial attention map, where $x_{i,j}$ encodes the dependence between the $i$-th location and $j$-th location. Then we input $\textbf{X}_k$ into a softmax layer and perform a matrix multiplication between the $\textbf{X}_k$ and $\textbf{D}_k$
\begin{equation}
\label{equa6}
\textbf{U}_k=softmax(\textbf{X}_k)\textbf{D}_k
\end{equation}
where $\textbf{U}_k\in R^{n\times C/r}$ and we reshape and transpose it to $R^{h\times w\times C/r}$. Then the $\textbf{U}_k$ is feeded to a $1\times1$ convolutional layer $\phi(\cdot)$ with output channel $=C$ and the residual connection with original feature $\textbf{F}_k$ is adopted.
\begin{equation}
\label{equa7}
\textbf{S}_k=\phi(\textbf{U}_k)+\textbf{F}_k
\end{equation}
After the whole sub-feature maps $\textbf{F}_k$ are refined, we obtain new feature map \textbf{S} containing rich spatial contextual information. Finally, we append such PSNL module on the tail of our proposed AWAN (see in Fig. \ref{figure2}).

\subsection{Camera Spectral Sensitivity (CSS) Prior}
\label{Camera Spectral Sensitivity (CSS) Prior}
Previous existing CNN-based SR models invariably to fit a brute-force RGB-to-HSI mapping and hardly consider to integrate camera spectral sensitivity (CSS) prior into SR for more accurate reconstruction. Based on the fact that the recovered RGB images can be created through applying the given CSS function to reconstructed HSI, we incorporate the discrepancies of the RGB images and differences of HSIs as a finer constraint. Accordingly, our loss function is a linear combination of two terms:
\begin{equation}
\label{equa8}
l = l_{h} + \tau l_{r}
\end{equation}
where ${\tau}$ is the tradeoff parameter. Given the ground truth ${{\textbf{I}}_{HSI}}$ and the spectral reconstructed HSI ${{\textbf{I}}_{SR}}$, the two loss functions are specifically defined as
\begin{equation}
\label{equa9}
l_{h}=\frac{1}{N}\sum_{p=1}^{N}(|\textbf{I}_{HSI}^{(p)}-\textbf{I}_{SR}^{(p)}|/\textbf{I}_{HSI}^{(p)}) 
\end{equation}
\begin{equation}
\label{equa10}
l_{r}=\frac{1}{N}\sum_{p=1}^{N}(|\mathbf{\Phi}(\textbf{I}_{HSI}^{(p)})-\mathbf{\Phi}(\textbf{I}_{SR}^{(p)})|)) 
\end{equation}
where ${\textbf{I}_{HSI}^{(p)}}$ and  ${\textbf{I}_{SR}^{(p)}}$ denote the ${p}$-th pixel value and ${\mathbf{\Phi}}$ is CSS function. ${N}$ is the total number of pixels. In our experiments, ${\tau}$ is set to 10 empirically.

\section{Experiments}
\subsection{Settings}
\label{Settings}

${\textbf{Hyperspectral datasets.}}$ In this paper, we evaluate our AWAN network on two challenging spectral reconstruction challenge datasets: NTIRE2018 \cite{arad2018ntire} and NTIRE2020 \cite{Arad_2020_CVPR_Workshops}. Both of two challenges are divided into two tracks: ``Clean'' and ``Real World''. The ``Clean'' track aims to recover HSIs from the noise-free RGB images obtained by a known CSS function, while the ``Real World'' track requires participants to rebuild the HSIs from JPEG-compression RGB images created by the an unknown camera response function. Note that the CSS functions of the same tracks are also different. Thus, there are four established benchmarks in total for SR in these two challenges. The NTIRE2018 dataset contains 256 natural HSIs for training and 5 + 10 additional images for validation and testing. All images are ${1392 \times 1300}$ in spatial size and have 31 spectral bands (400-700nm at roughly 10nm increments). The NTIRE2020 dataset consists of 450 images for training, 10 images for validation and 20 images for testing with a spatial resolution of ${512 \times 482}$. The band number is also 31.

\begin{table*}[]
	\vskip 0.05in
	\centering
	\begin{tabular}{|l|c|c|c|c|c|c|c|c|c|}
		\hline
		& \multicolumn{5}{c|}{Clean}                 & \multicolumn{4}{c|}{Real World}   \\ \hline
		Description &$E_a$&$E_b$&$E_c$&$E_d$&$E_e$&$E_f$&$E_g$&$E_h$&$E_i$\\ \hline
		PSNL &\ding{56}&\ding{52}&\ding{56}&\ding{52}&\ding{52}&\ding{56}&\ding{52}&\ding{56}&\ding{52}\\ \hline
		AWAC&\ding{56}&\ding{56}&\ding{52}&\ding{52}&\ding{52}&\ding{56}&\ding{56}&\ding{52}&\ding{52}\\ \hline
		CSS&\ding{56}&\ding{56}&\ding{56}&\ding{56}&\ding{52}&\ding{56}&\ding{56}&\ding{56}&\ding{56}\\ \hline
		MRAE
		& 0.0359 & 0.0350 & 0.0341 & 0.0326 & 0.0321 & 0.0687 & 0.0679 & 0.0672 & 0.0668 \\ \hline
	\end{tabular}
	\caption{Ablation study on validation set of NTIRE2020 ``Clean'' and ``Real World'' tracks. We report the best MRAE values in $3 \times 10^5$ iterations.}
	\label{table1}
	\vskip -0.14in
\end{table*}

${\textbf{Evaluation metrics.}}$ To objectively evaluate the performance of our proposed method on the NTIRE2020 and NTIRE2018 datasets, the root mean square error (RMSE) and mean relative absolute error (MRAE) are utilized as evaluation metrics following the scoring script provided by the challenge. MRAE is chosen as the ranking criterion rather than RMSE to avoid overweighting errors in higher luminance areas of the test images. MRAE and RMSE are calculated as follows
\begin{equation}
\label{equa11}
MRAE=\frac{1}{N} \sum_{p=1}^{N}\left(\left|\textbf{I}_{HSI}^{(p)}-\textbf{I}_{SR}^{(p)}\right| / \textbf{I}_{HSI}^{(p)}\right)
\end{equation}
\begin{equation}
\label{equa12}
RMSE=\sqrt{\frac{1}{N} \sum_{p=1}^{N}\left(\textbf{I}_{HSI}^{(p)}-\textbf{I}_{SR}^{(p)}\right)^{2}}
\end{equation}
where ${\textbf{I}_{HSI}^{(p)}}$ and ${\textbf{I}_{SR}^{(p)}}$ denote the ${p}$-th pixel value of the ground truth and the spectral reconstructed HSI. A smaller MRAE or RMSE indicate better performance.

${\textbf{Implementations details.}}$ We design DRAB number as ${M=8}$ and output channel $=200$. During the training process, we set up ${64 \times 64}$ RGB and HSI sample pairs from the original dataset. The batch size of our model is 32 and the parameter optimization algorithm chooses Adam \cite{kingma2014adam} with ${\beta_1 = 0.9}$, ${\beta_2 = 0.99}$ and ${\epsilon = 10^{-8}}$. The reduction ratio $t$ value of the AWCA module is 16 and $r$ value of PSNL module is 8. The learning rate is initialized as 0.0001 and the polynomial function is set as the decay policy with power $=1.5$. We stop network training at 100 epochs. Our proposed AWAN network has been implemented on the Pytorch framework and the training time is approximately 36 hours on 2 NVIDIA 2080Ti GPUs.

\begin{table}[]
	\vskip 0.05in
	\centering
	\begin{tabular}{|l|c|c|c|c|}
		\hline
		\multirow{2}{*}{Method} & \multicolumn{2}{c|}{Clean} & \multicolumn{2}{c|}{Real World} \\ \cline{2-5} 
		&MRAE & RMSE& MRAE& RMSE\\ \hline
		AWAN+                & \textbf{0.0312}           & \textbf{0.0111}          & \textbf{0.0639}              & \textbf{0.0170}             \\ \hline
		AWAN              & \underline{0.0321}           & \underline{0.0112}           & \underline{0.0668}              & \underline{0.0175}             \\ \hline
		HSCNN-R~\cite{shi2018hscnn+}              & 0.0372           & 0.0143           & 0.0684              & 0.0182             \\ \hline
		Stiebel~\cite{stiebel2018reconstructing}              & 0.0395           & 0.0152           & 0.0698              & 0.0187             \\ \hline
		Yan~\cite{yan2018accurate}              	 & 0.0724           & 0.0235           & 0.0875              & 0.0225             \\ \hline
		Galliani~\cite{galliani2017learned}             & 0.0850           & 0.0251           & 0.0941              & 0.0243             \\ \hline
		Arad~\cite{arad2016sparse}                 & 0.0787           & 0.0331           & ------              & ------             \\ \hline
	\end{tabular}
	\caption{The quantitative results on validation set of NTIRE2020 ``Clean'' and ``Real World'' tracks. The best and second best results are \textbf{highlighted} and \underline{underlined}.}
	\label{table2}
	\vskip -0.14in
\end{table}

\begin{table}[]
	\vskip 0.05in
	\centering
	\begin{tabular}{|l|c|c|c|c|}
		\hline
		\multirow{2}{*}{Method} & \multicolumn{2}{c|}{Clean} & \multicolumn{2}{c|}{Real World} \\ \cline{2-5} 
		&MRAE & RMSE& MRAE& RMSE\\ \hline
		AWAN+                &\textbf{0.0114}   &\textbf{10.24}   &\textbf{0.0277}      &\textbf{21.33}   \\ \hline
		AWAN                 &\underline{0.0116}   &\underline{10.49}   &0.0289      &\underline{22.18}   \\ \hline
		HSCNN-D\cite{shi2018hscnn+}              & 0.0131           & 12.99           & \underline{0.0288}              & 22.40             \\ \hline
		HSCNN-R\cite{shi2018hscnn+}              & 0.0134           & 13.17           & 0.0297              & 22.88             \\ \hline
		Stiebel\cite{stiebel2018reconstructing}              & 0.0156           & 15.88           & 0.0312              & 23.88             \\ \hline
		Yan\cite{yan2018accurate}              	 & 0.0231           & 19.28           & 0.0389              & 31.65             \\ \hline
		Galliani\cite{galliani2017learned}             & 0.0272           & 20.98           & 0.0660              & 55.19             \\ \hline
		Arad\cite{arad2016sparse}                 & 0.0808           & 51.48           & ------              & ------             \\ \hline
	\end{tabular}
	\caption{The quantitative results of validation set of NTIRE2018 ``Clean'' and ``Real World'' tracks. The best and second best results are \textbf{highlighted} and \underline{underlined}.}
	\label{table3}
	\vskip -0.14in
\end{table}

\subsection{Ablation Analysis}
\label{Ablation Analysis}
To verify the effects of different modules, we carry out ablation study on the NTIRE2020 ``Clean" and ``Real World" tracks. The detailed experimental results are listed in the TABLE \ref{table1}. $E_a$ and $E_f$ refer to a baseline network stacked with $8$ DRABs, which only contains plenty of ordinary convolutional layers.

\begin{figure*}[!tbp]
	\vskip 0.05in
	\centering
	\scalebox{1}
	{
		\begin{tabular}{@{}c@{}c@{}c@{}c@{}c@{}c@{}c@{}c}
			& $\text{Galliani~\cite{galliani2017learned}}$ &$\text{Yan~\cite{yan2018accurate}}$ &$\text{Stiebel~\cite{stiebel2018reconstructing}}$&$\text{HSCNN-R~\cite{shi2018hscnn+}}$&$\text{Ours}$ \\
			&{\includegraphics[width=2.41cm,height=2.41cm]{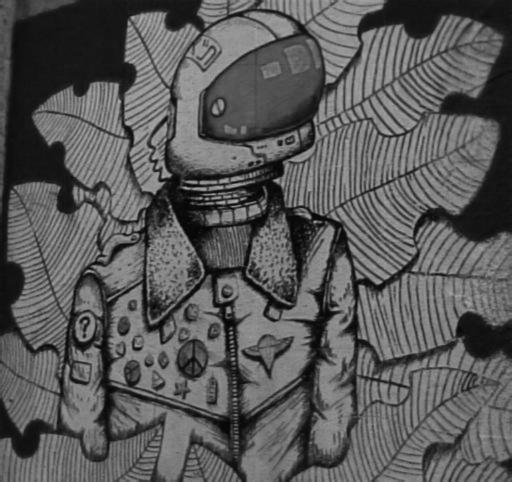}} \
			&{\includegraphics[width=2.41cm,height=2.41cm]{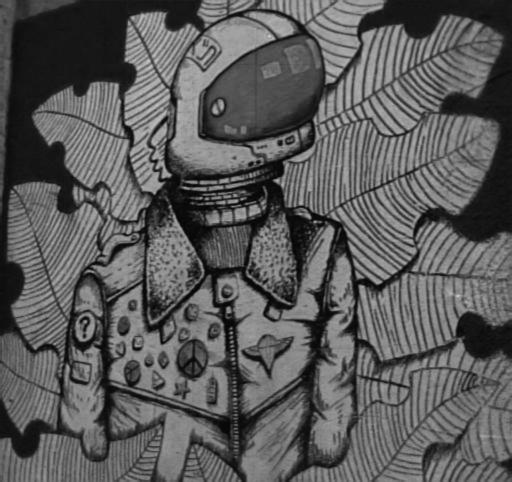}} \
			&{\includegraphics[width=2.41cm,height=2.41cm]{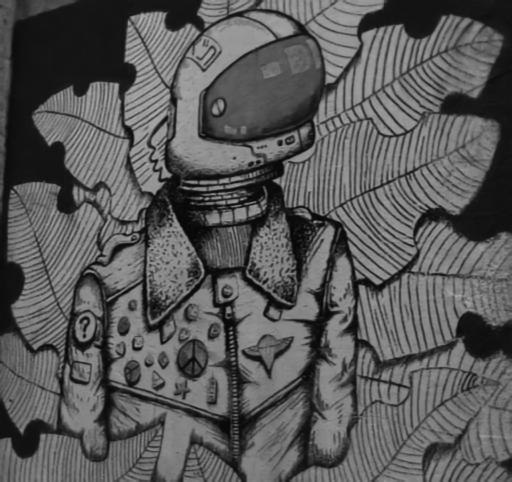}} \
			&{\includegraphics[width=2.41cm,height=2.41cm]{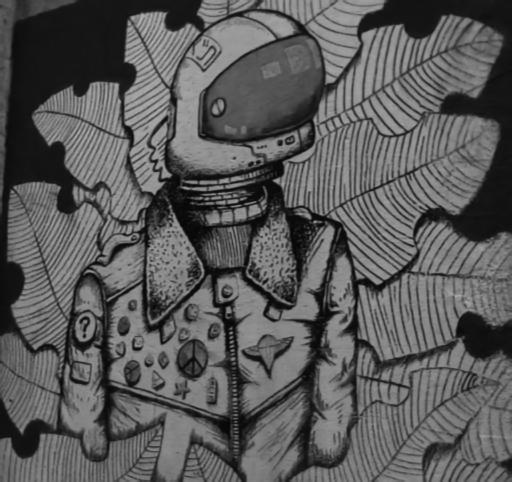}} \
			&{\includegraphics[width=2.41cm,height=2.41cm]{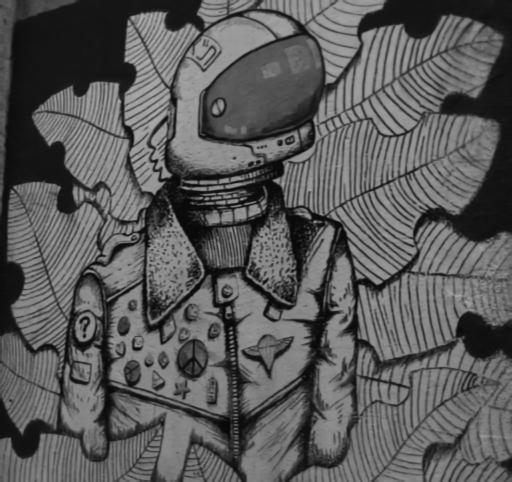}} \\
			&{\includegraphics[width=2.41cm,height=2.41cm]{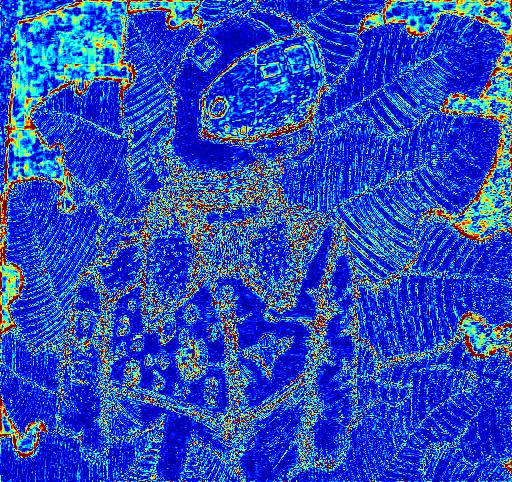}} \
			&{\includegraphics[width=2.41cm,height=2.41cm]{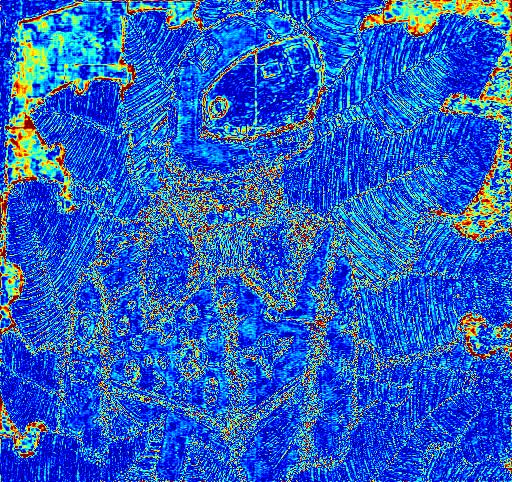}} \
			&{\includegraphics[width=2.41cm,height=2.41cm]{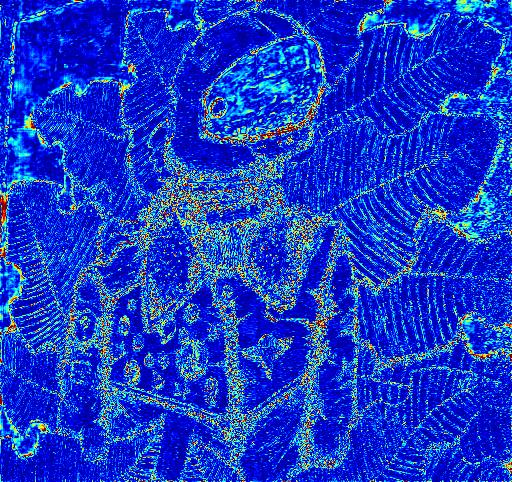}} \
			&{\includegraphics[width=2.41cm,height=2.41cm]{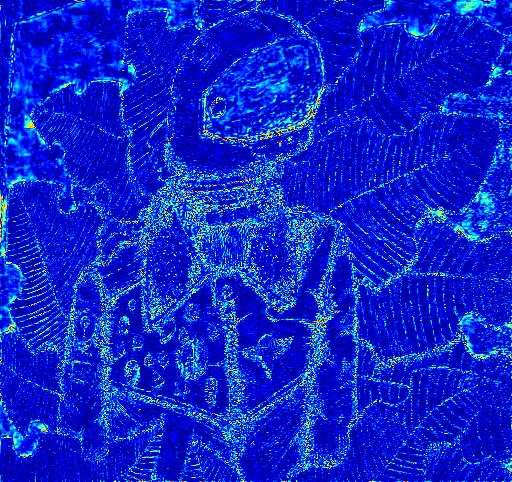}} \
			&{\includegraphics[width=2.41cm,height=2.41cm]{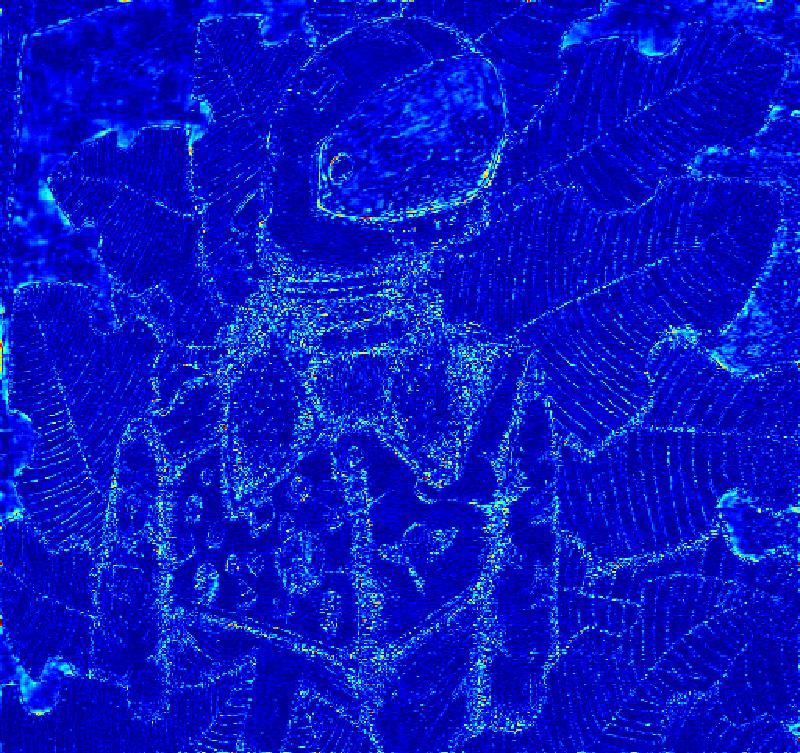}}\
			&{\includegraphics[width=0.2cm,height=2.41cm]{figure_prior_455/colorbar3.png}} 
			\\
		\end{tabular}
	}
	\caption{The Visual results of the 22-th band and the reconstruction error images of an HSI chosen from validation set of NTIRE2020 ``Real World'' track. The error images are the heat maps of MRAE between the ground truth and the recovered HSI. The best view on the screen.}
	\label{figure5}	
	\vskip -0.14in
\end{figure*}

\begin{figure*}[!tbp]
	\vskip 0.05in
	\centering
	\scalebox{1}
	{
		\begin{tabular}{@{}c@{}c@{}c@{}c@{}c@{}c@{}c@{}c}
			&$\text{Galliani~\cite{galliani2017learned}}$ &$\text{Yan~\cite{yan2018accurate}}$ &$\text{Stiebel~\cite{stiebel2018reconstructing}}$
			&$\text{HSCNN-R~\cite{shi2018hscnn+}}$
			&$\text{HSCNN-D~\cite{shi2018hscnn+}}$
			&$\text{Ours}$ \\
			&{\includegraphics[width=2.41cm,height=2.41cm]{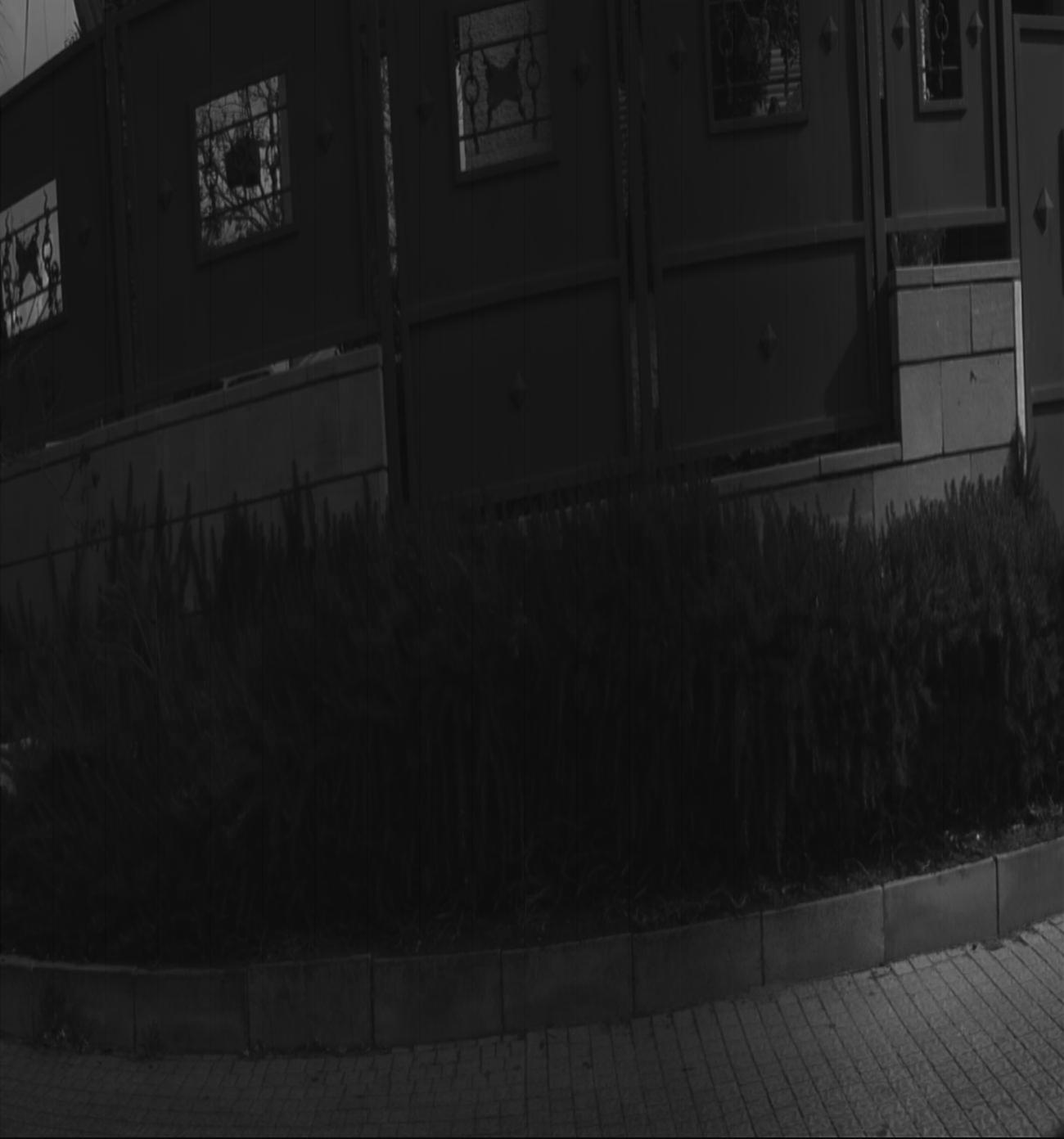}} \
			&{\includegraphics[width=2.41cm,height=2.41cm]{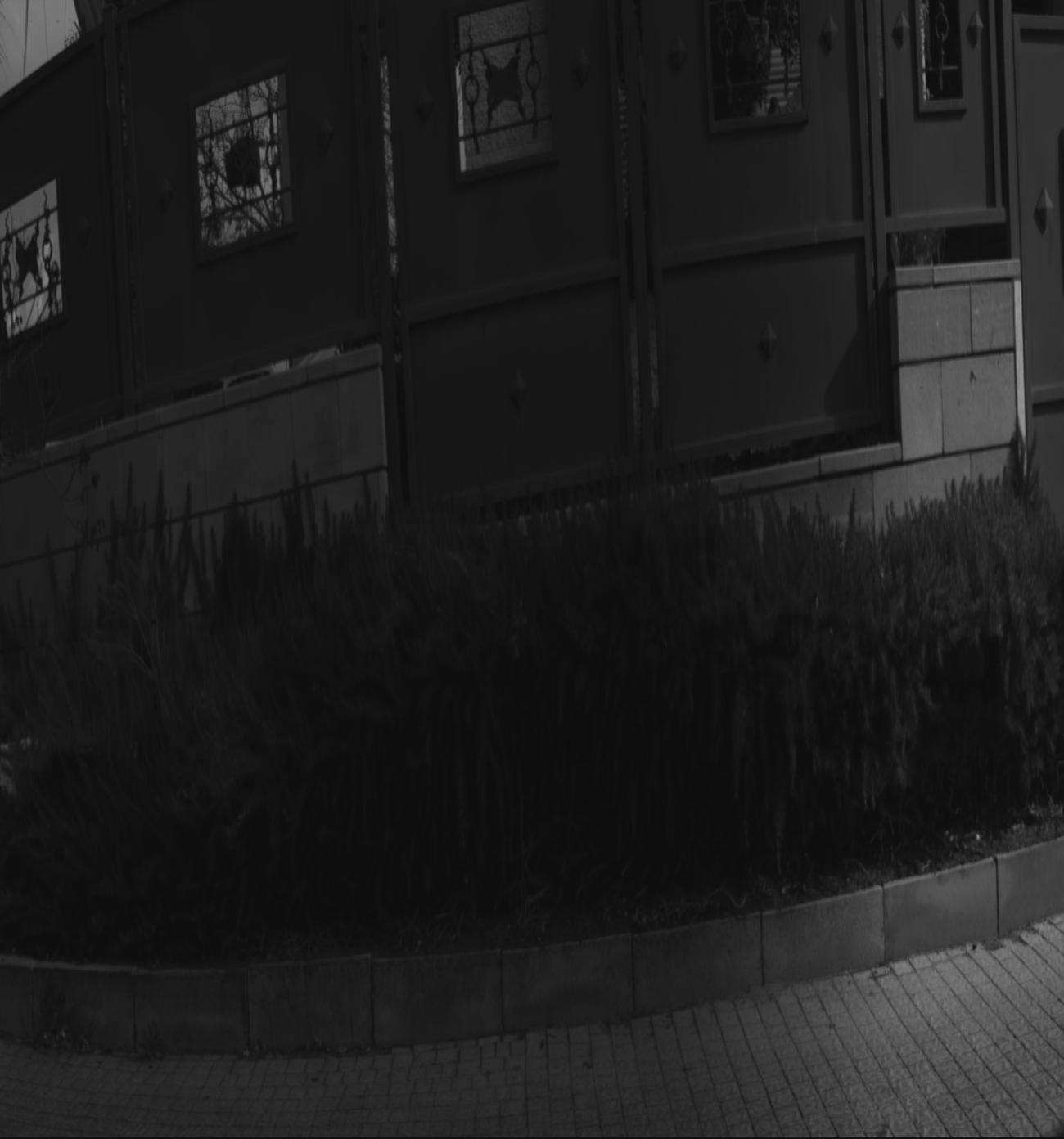}} \
			&{\includegraphics[width=2.41cm,height=2.41cm]{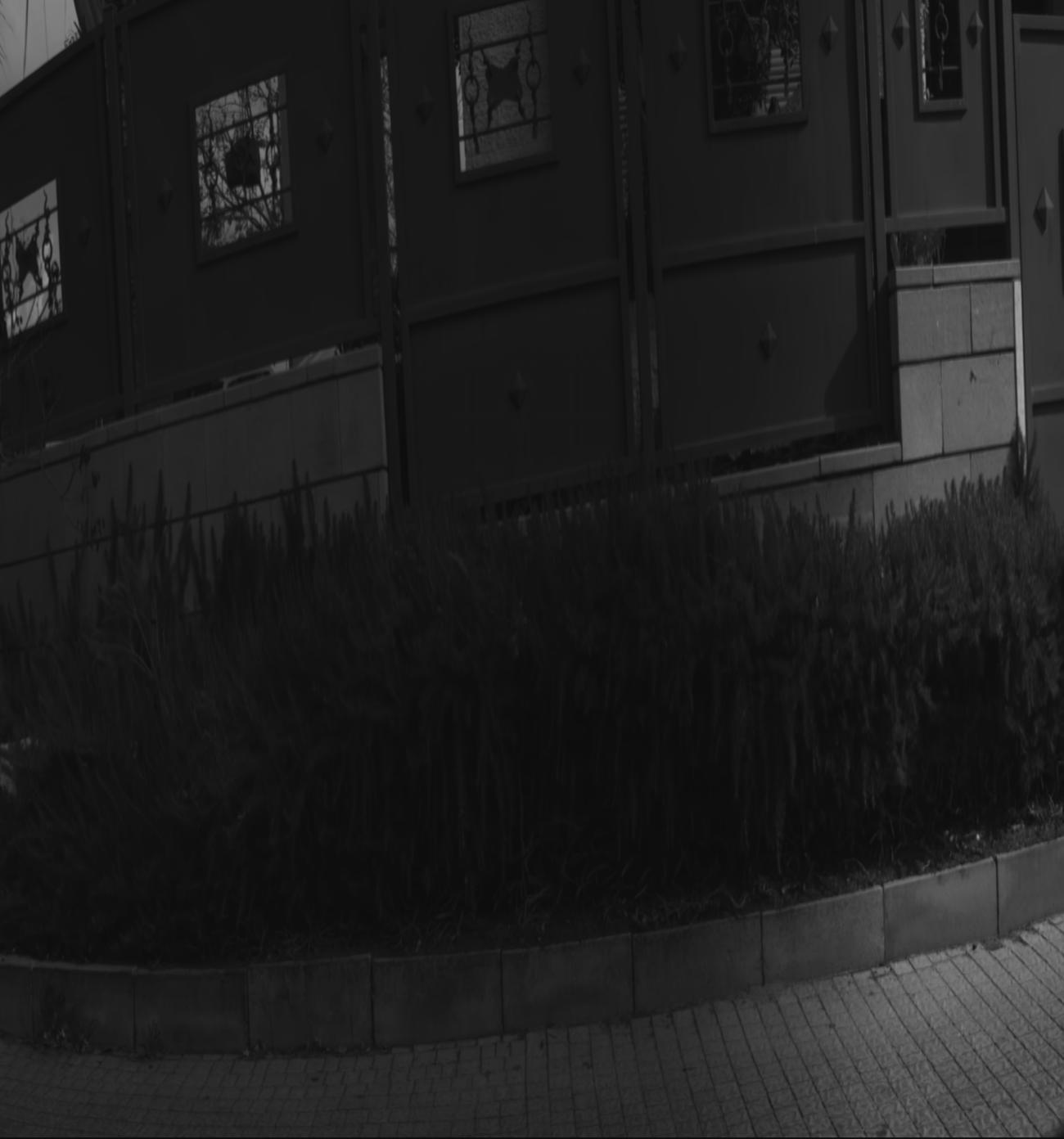}} \
			&{\includegraphics[width=2.41cm,height=2.41cm]{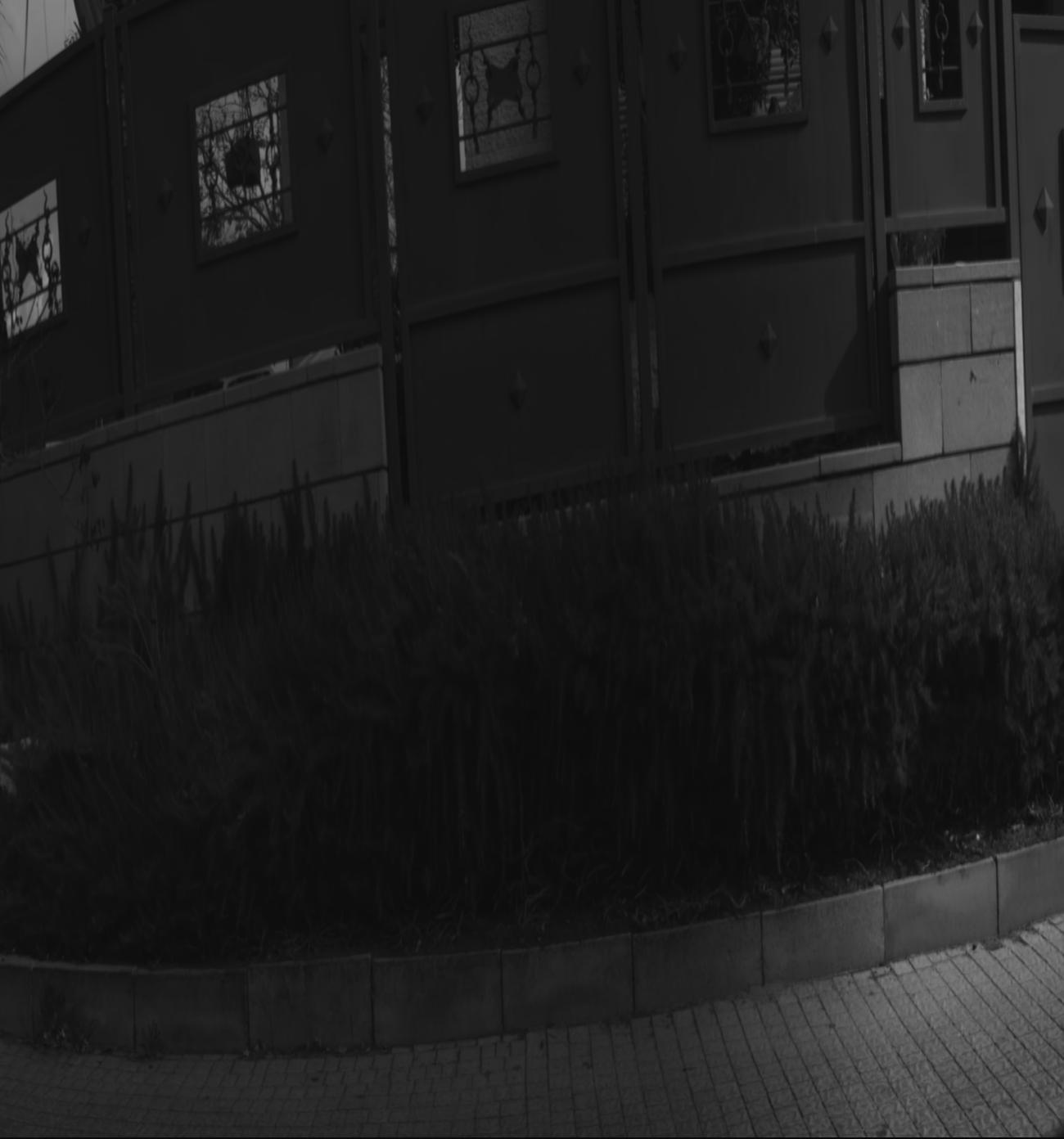}} \
			&{\includegraphics[width=2.41cm,height=2.41cm]{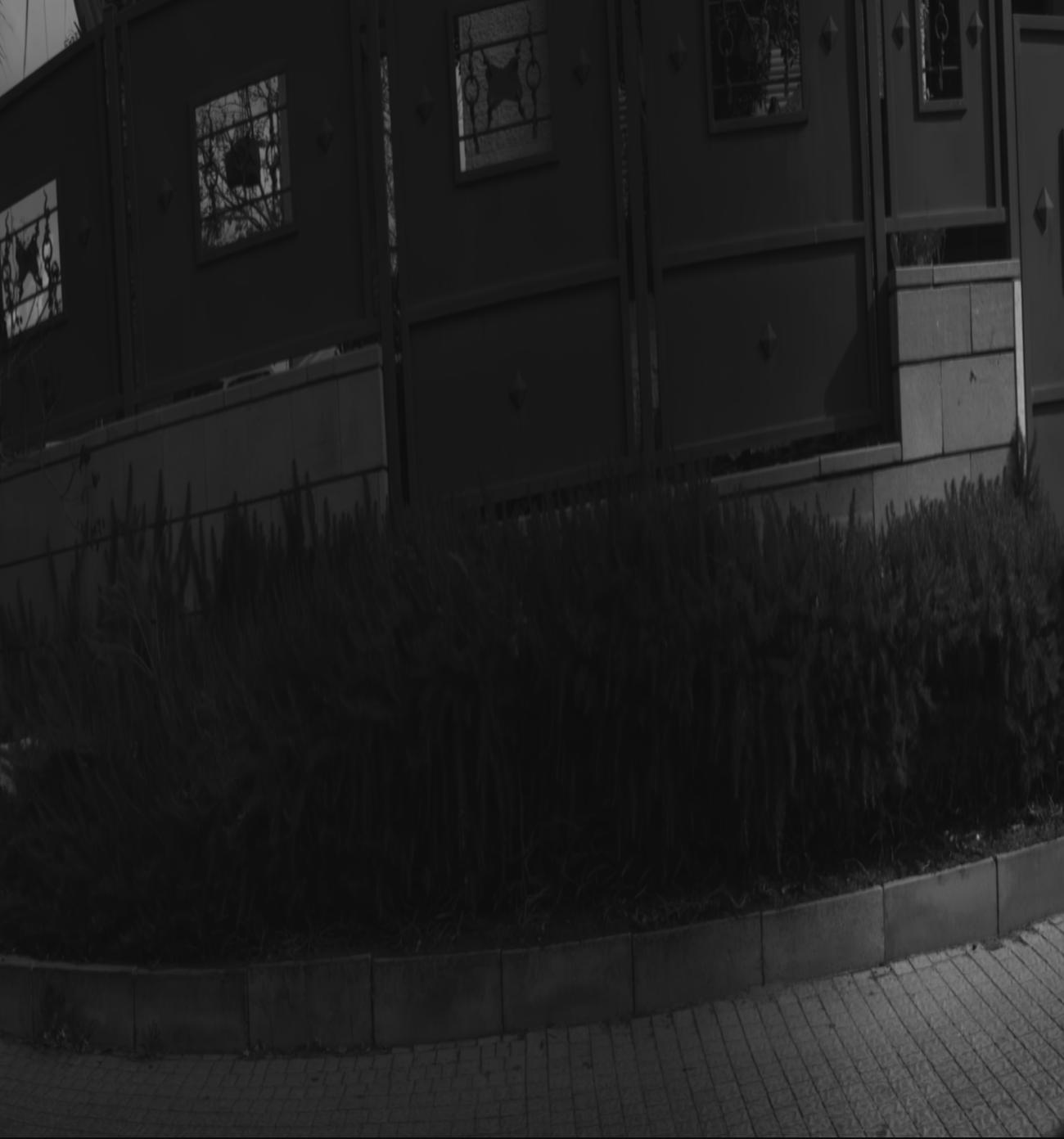}} \
			&{\includegraphics[width=2.41cm,height=2.41cm]{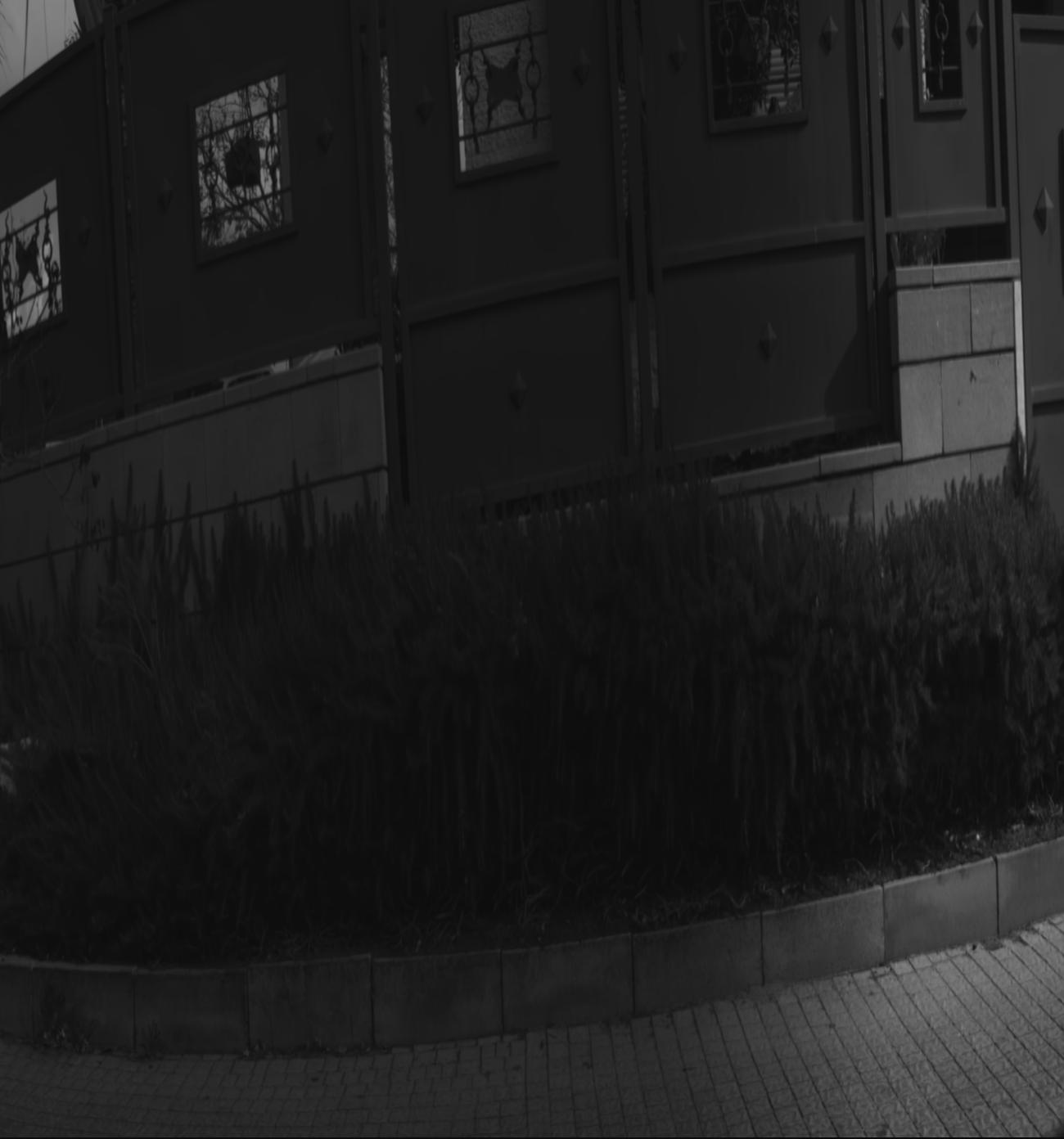}} \\
			&{\includegraphics[width=2.41cm,height=2.41cm]{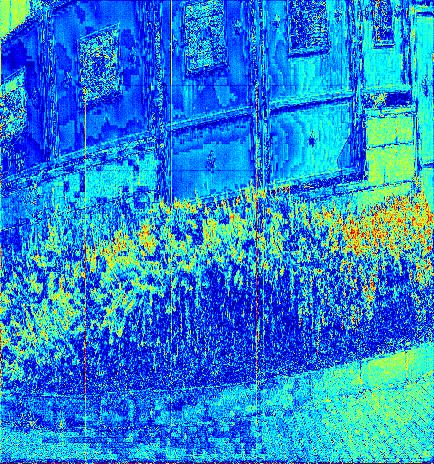}} \
			&{\includegraphics[width=2.41cm,height=2.41cm]{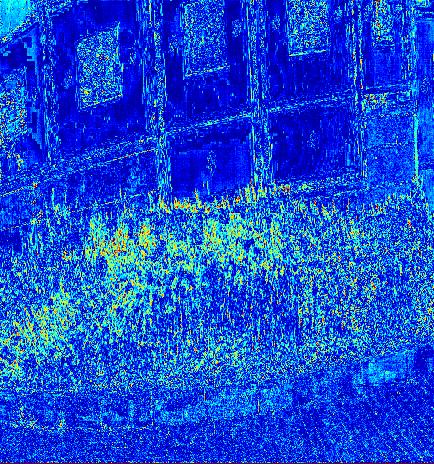}} \
			&{\includegraphics[width=2.41cm,height=2.41cm]{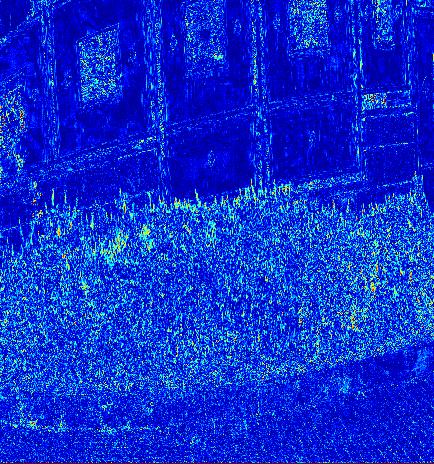}} \
			&{\includegraphics[width=2.41cm,height=2.41cm]{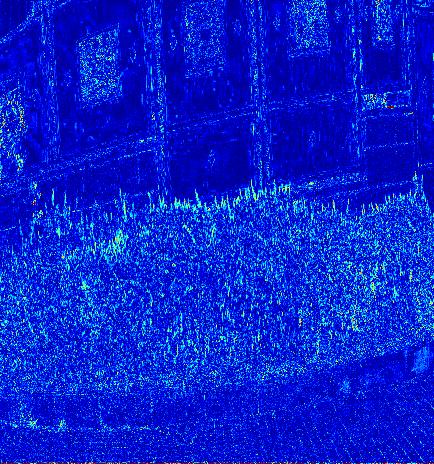}} \
			&{\includegraphics[width=2.41cm,height=2.41cm]{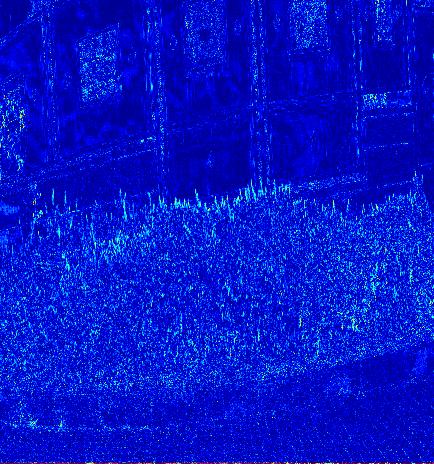}} \
			&{\includegraphics[width=2.41cm,height=2.41cm]{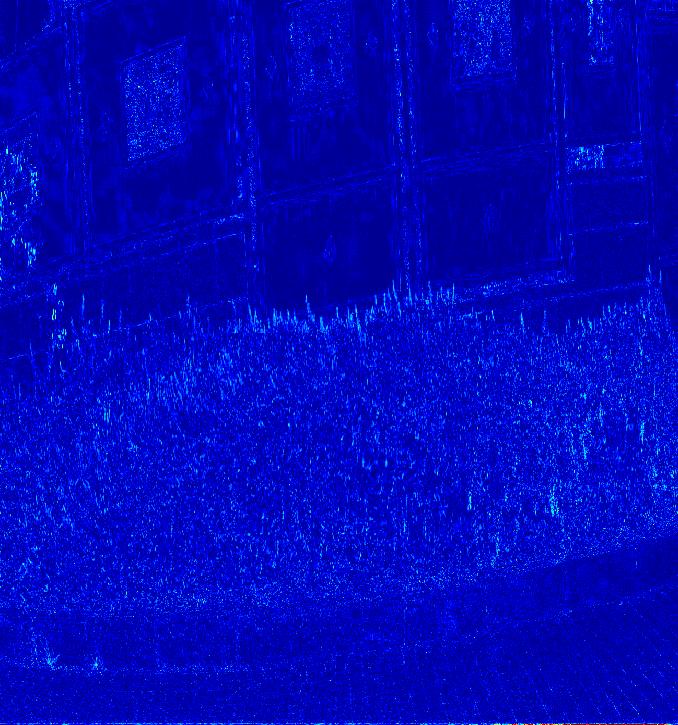}}\
			&{\includegraphics[width=0.2cm,height=2.41cm]{figure_prior_455/colorbar3.png}} 
			\\
		\end{tabular}
	}
	\caption{The Visual results of the 15-th band and the reconstruction error images of an HSI chosen from validation set of NTIRE2018 ``Real World'' track. The error images are the heat maps of MRAE between the ground truth and the recovered HSI. The best view on the screen.}
	\label{figure7}	
	\vskip -0.14in
\end{figure*}

\begin{figure*}[!tbp]
	\vskip 0.05in
	\centering
	\scalebox{1}
	{
		\begin{tabular}{@{}c@{}c@{}c@{}c@{}c@{}c@{}c@{}c@{}c}
			&$\text{Arad~\cite{arad2016sparse}}$ &${\text{Galliani~\cite{galliani2017learned}}}$ & $\text{Yan~\cite{yan2018accurate}}$ &$\text{Stiebel~\cite{stiebel2018reconstructing}}$ &$\text{HSCNN-R~\cite{shi2018hscnn+}}$&$\text{HSCNN-D~\cite{shi2018hscnn+}}$&$\text{Ours}$ \\
			&{\includegraphics[width=2.41cm,height=2.41cm]{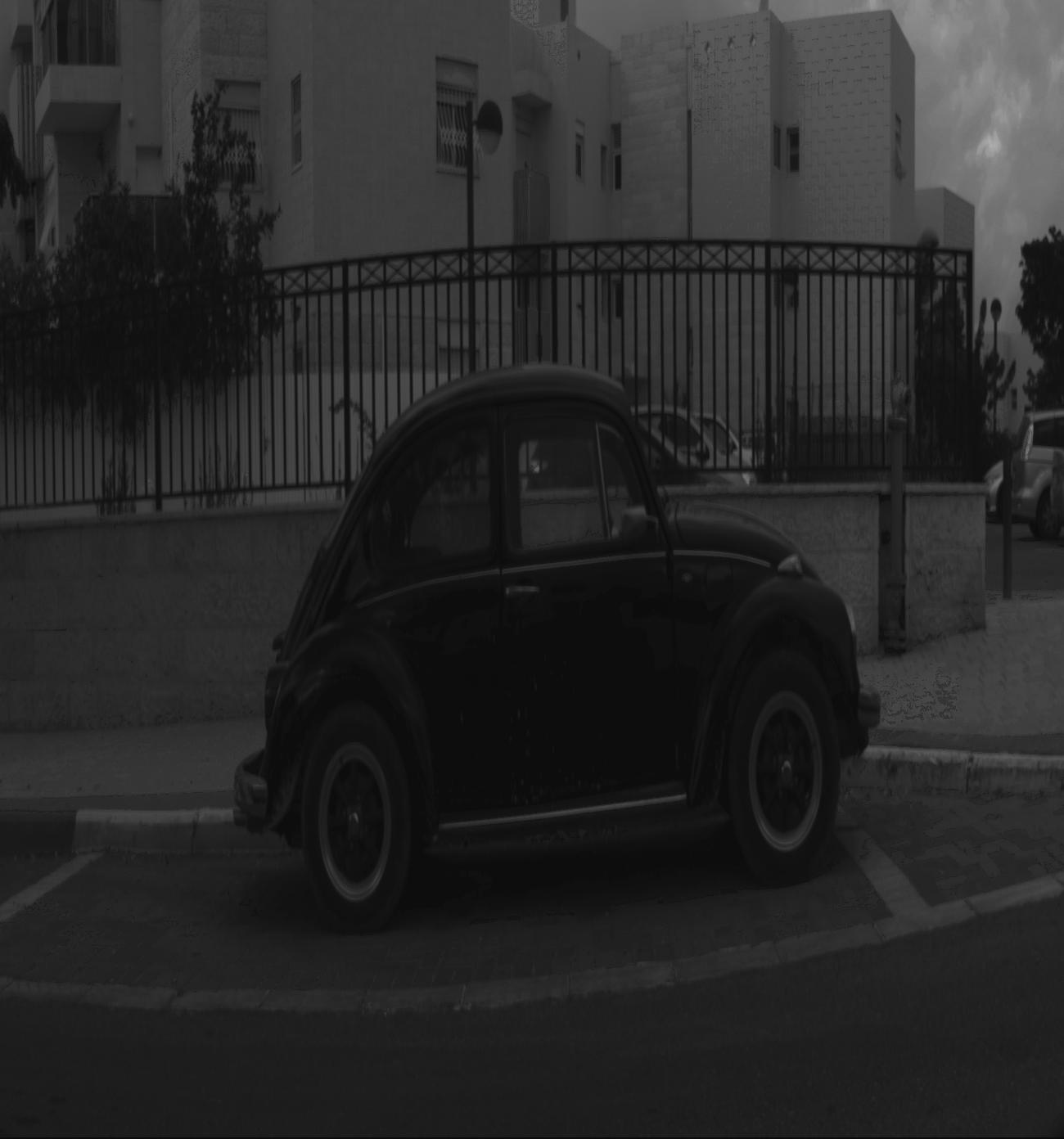}} \
			&{\includegraphics[width=2.41cm,height=2.41cm]{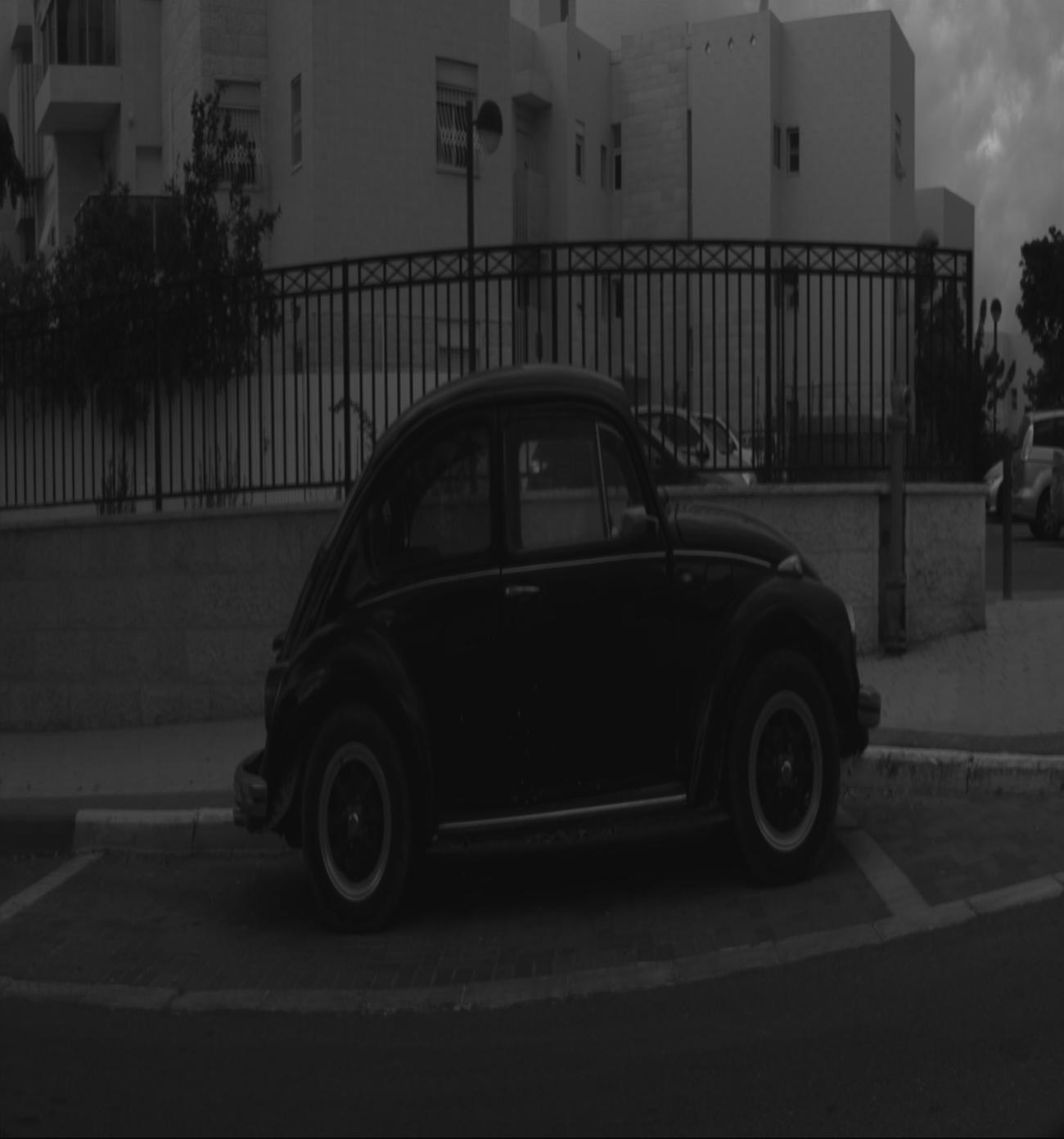}} \
			&{\includegraphics[width=2.41cm,height=2.41cm]{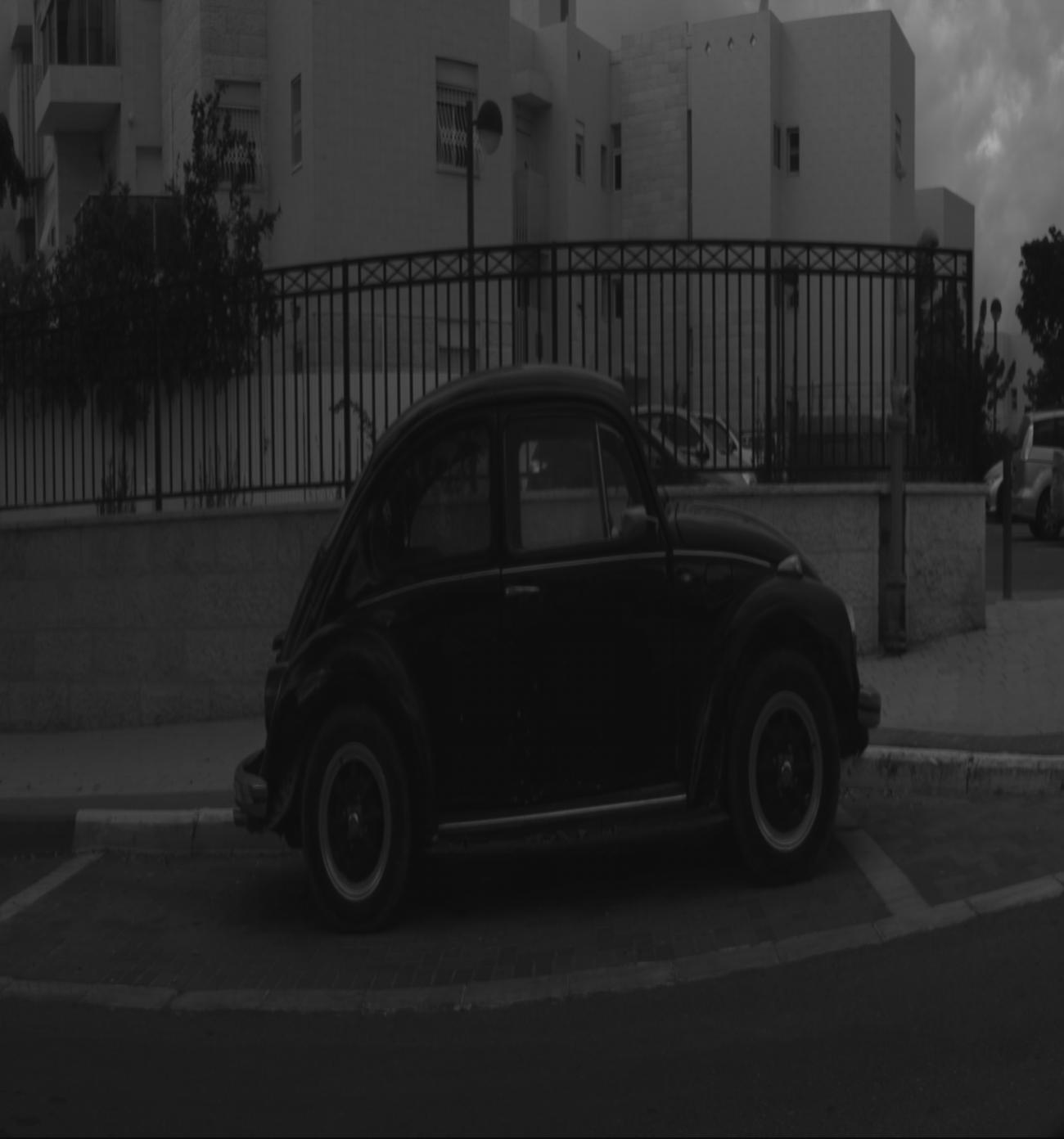}} \
			&{\includegraphics[width=2.41cm,height=2.41cm]{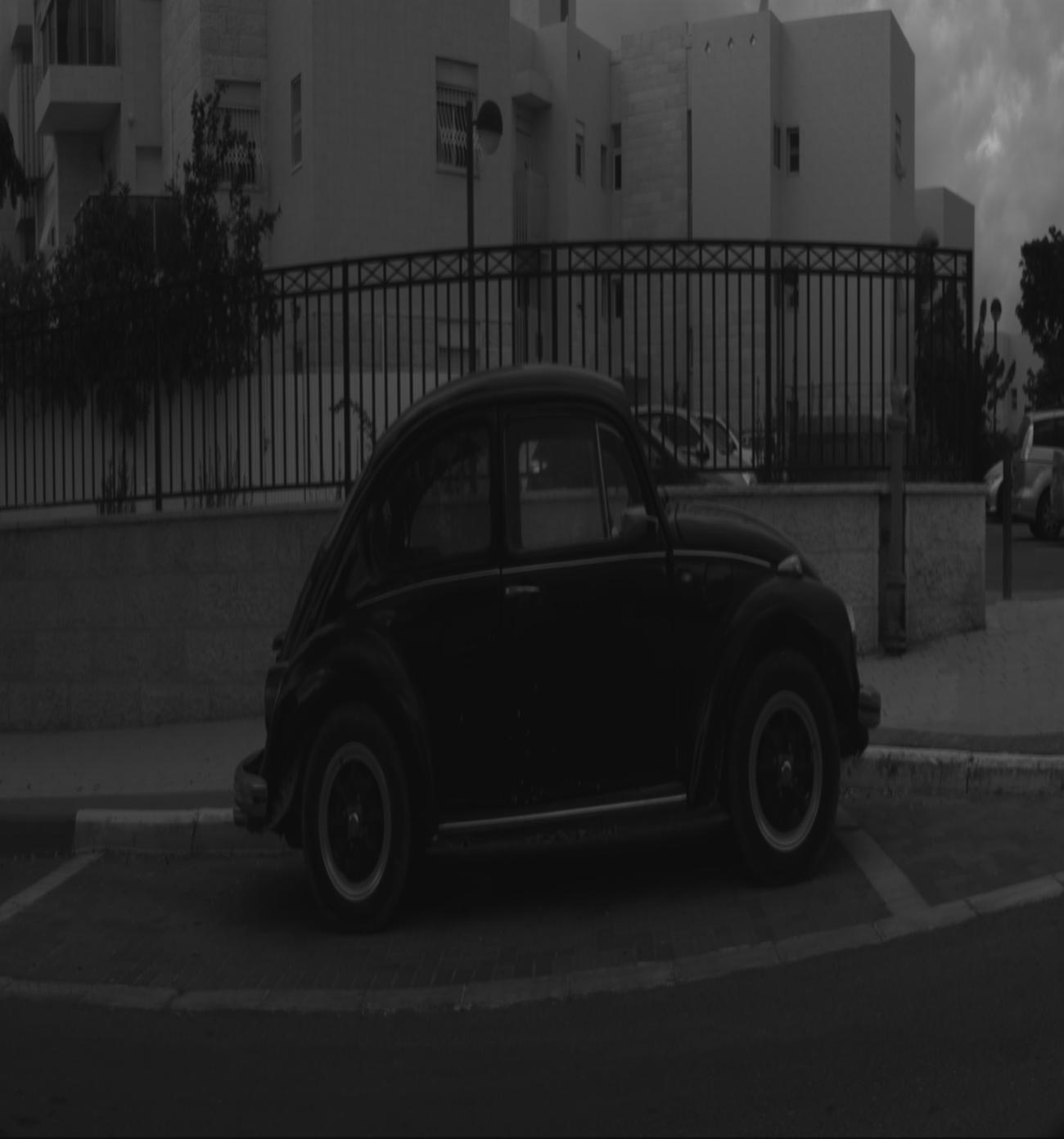}} \
			&{\includegraphics[width=2.41cm,height=2.41cm]{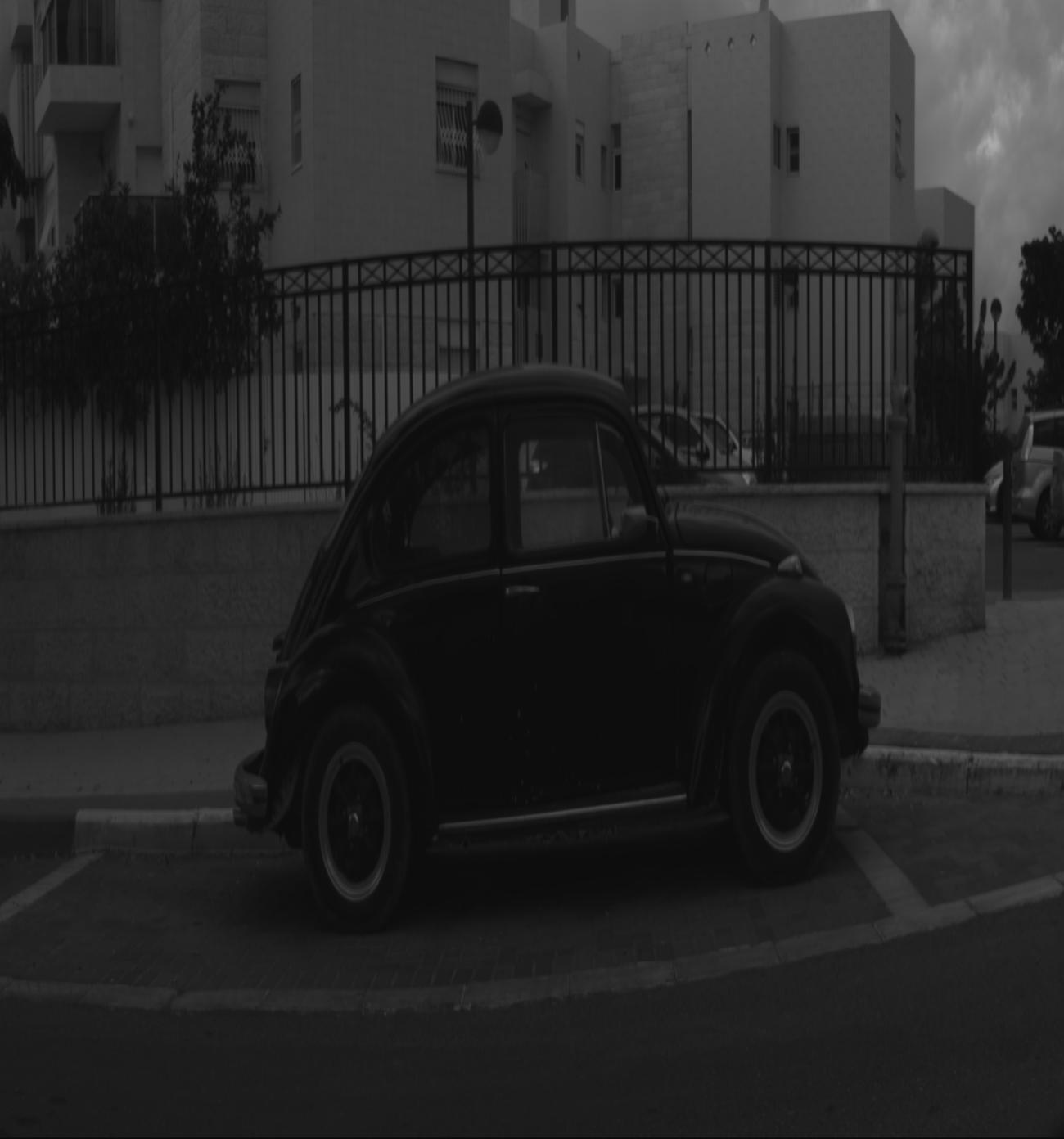}} \
			&{\includegraphics[width=2.41cm,height=2.41cm]{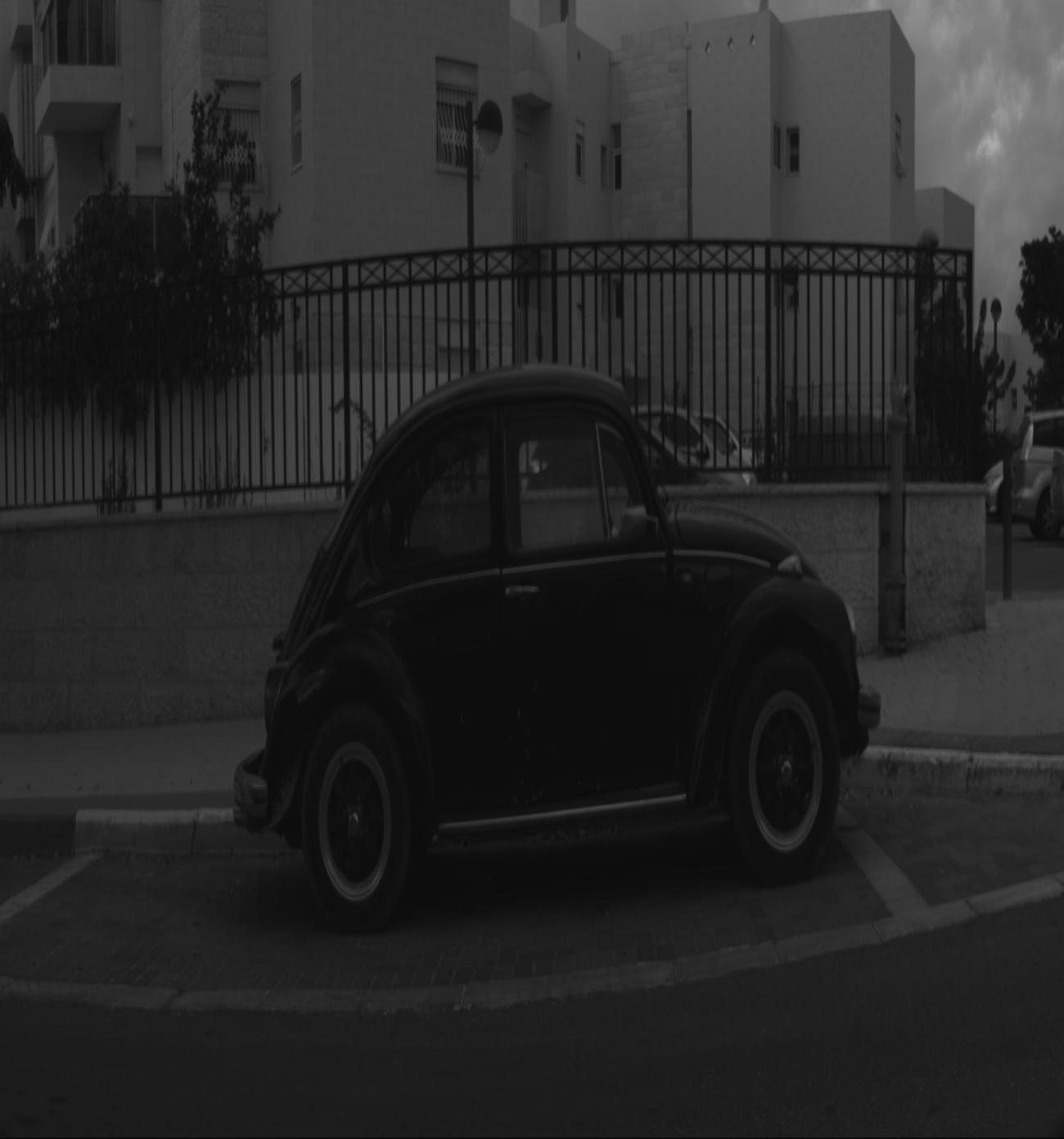}} \
			&{\includegraphics[width=2.41cm,height=2.41cm]{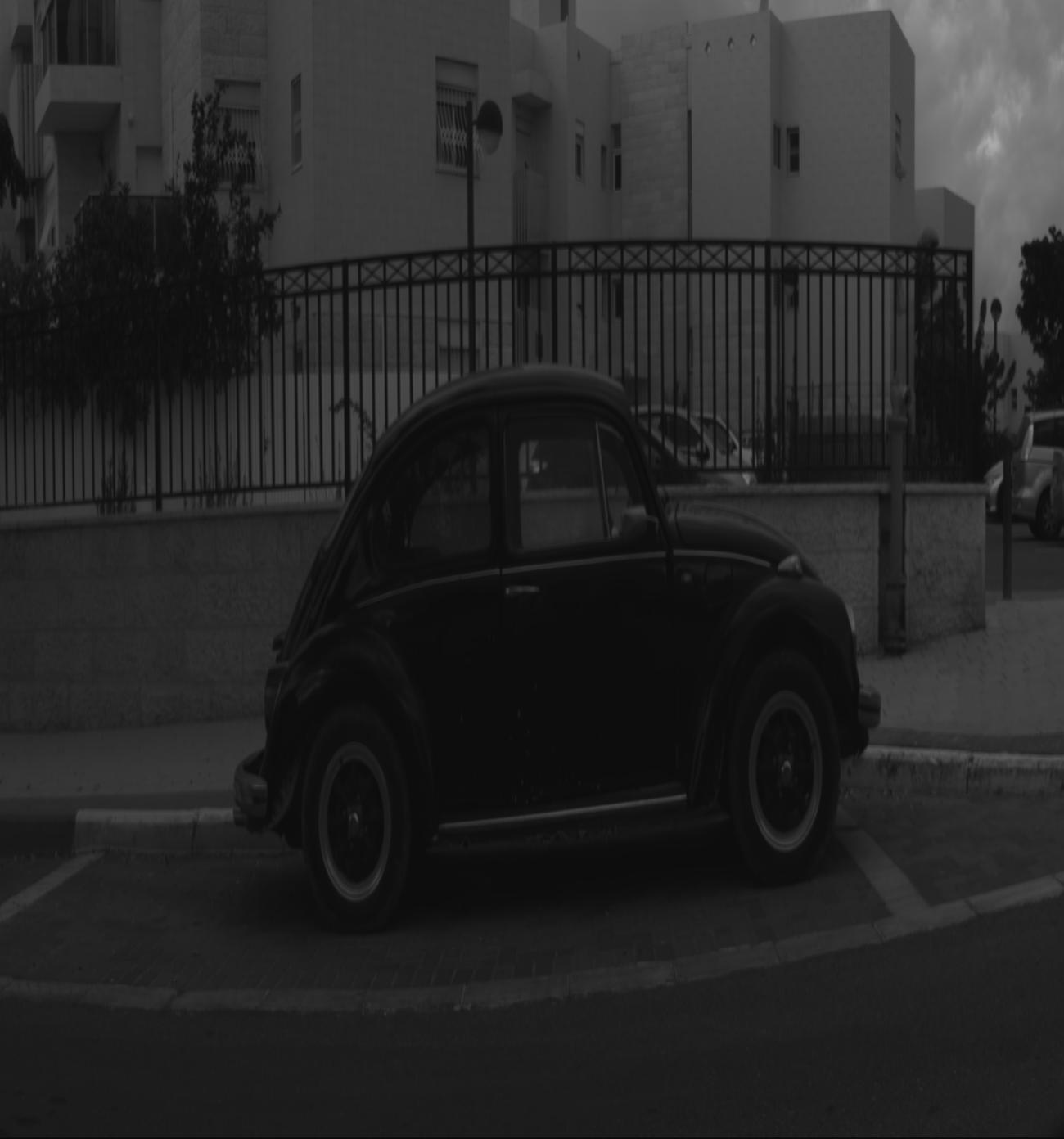}} \\
			&{\includegraphics[width=2.41cm,height=2.41cm]{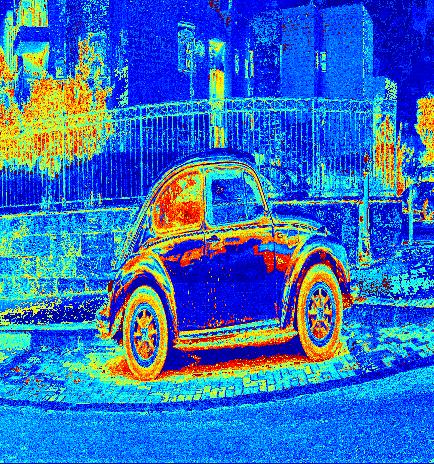}} \
			&{\includegraphics[width=2.41cm,height=2.41cm]{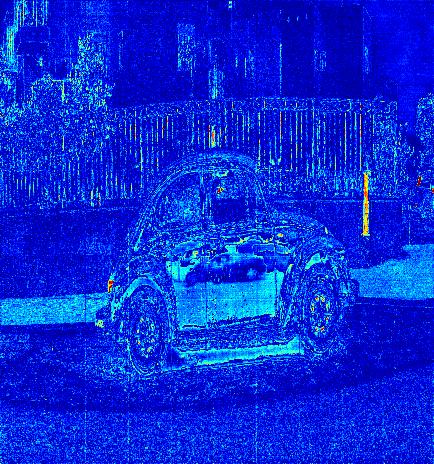}} \
			&{\includegraphics[width=2.41cm,height=2.41cm]{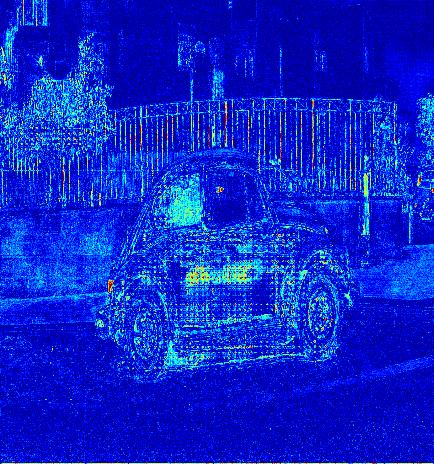}} \
			&{\includegraphics[width=2.41cm,height=2.41cm]{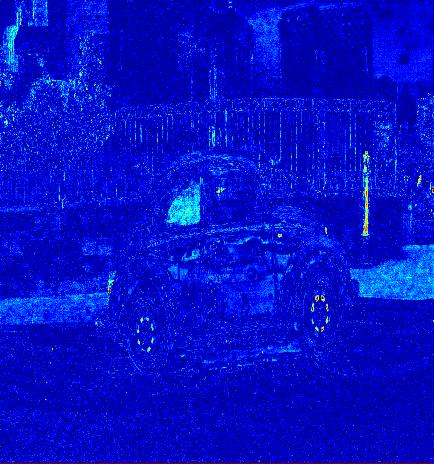}} \
			&{\includegraphics[width=2.41cm,height=2.41cm]{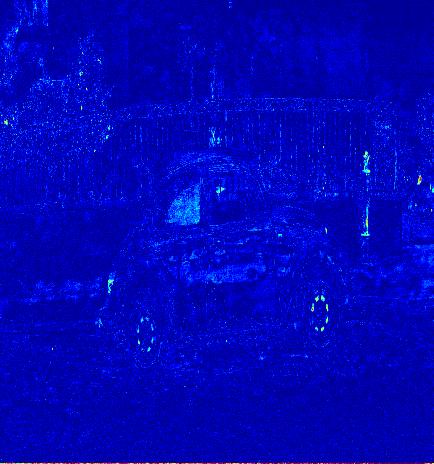}} \
			&{\includegraphics[width=2.41cm,height=2.41cm]{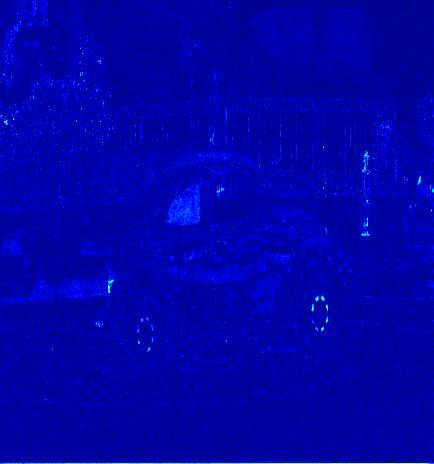}} \
			&{\includegraphics[width=2.41cm,height=2.41cm]{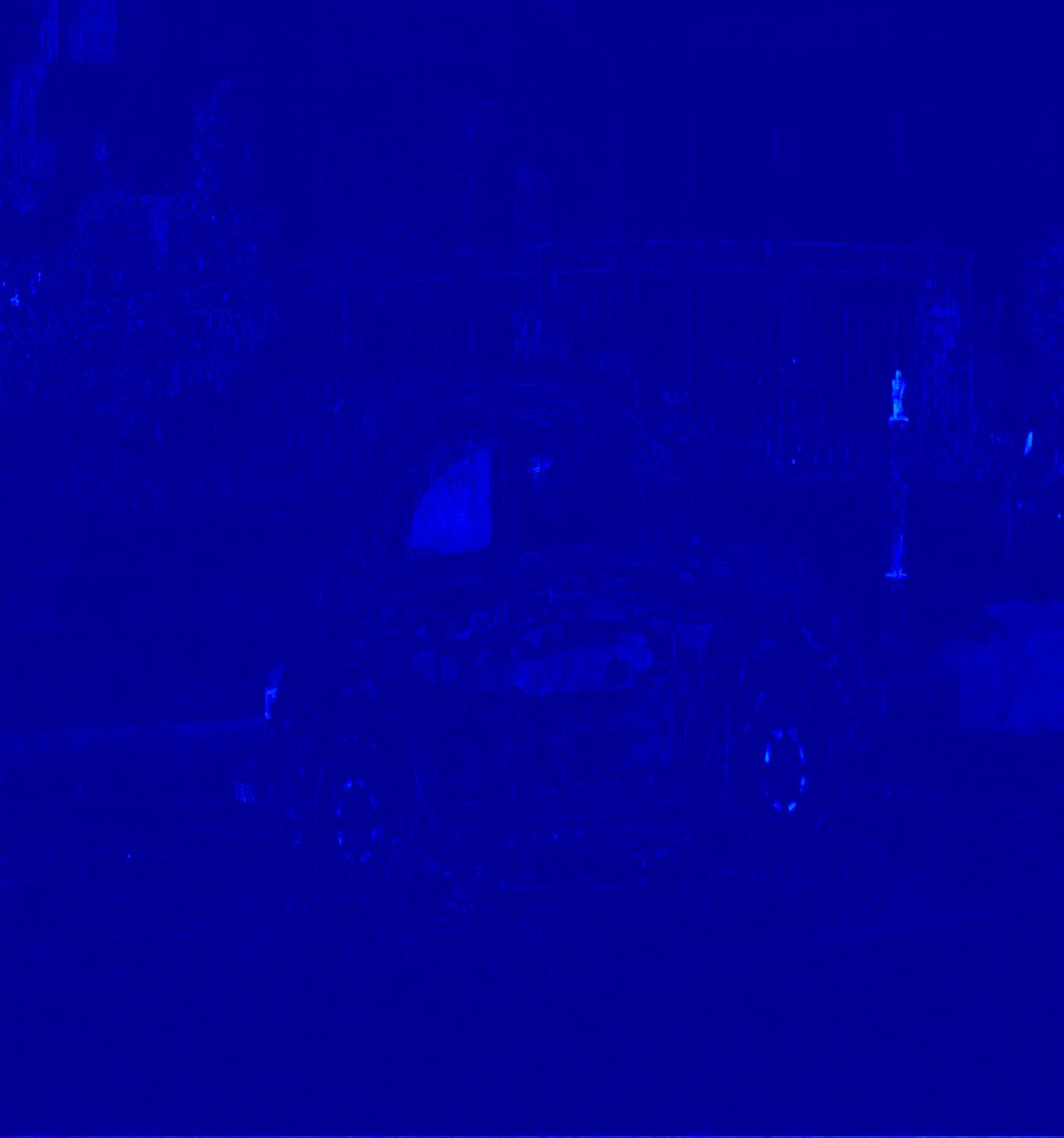}}\
			&{\includegraphics[width=0.2cm,height=2.41cm]{figure_prior_455/colorbar3.png}} 
			\\
		\end{tabular}
	}
	\caption{The Visual results of the 24-th band and the reconstruction error images of an HSI chosen from validation set of NTIRE2018 ``Clean'' track. The error images are the heat maps of MRAE between the ground truth and the recovered HSI. The best view on the screen.}
	\label{figure6}
	\vskip -0.05in	
\end{figure*}

\textbf{Patch-level second-order non-local (PSNL).} From TABLE \ref{table1} we can see that the baseline results reach MRAE=0.0359 and MRAE=0.0687 on the two tracks separately. As described in Section \ref{Patch-level Second-order Non-local (PSNL)}, we append the PSNL module on the tail of our proposed AWAN to capture the long-range dependencies via second-order non-local operations. Compared with the baseline results, $E_b$ and $E_g$ demonstrate the effectiveness of modeling distant region relationships.

\begin{table}[]
	\vskip 0.05in
	\centering
	\begin{tabular}{|l|l|l|l|}
		\hline
		Method    & MRAE   & Runtime/s & Compute Platform \\ \hline
		\textbf{AWAN+}     & \textbf{0.03010} & 0.56      & NVIDIA 2080Ti    \\ \hline
		2nd method   & 0.03076 & 16        & NVIDIA 1080Ti    \\ \hline
		3rd method    & 0.03231 & 3.748     & NVIDIA Titan XP  \\ \hline
		4th method   & 0.03476 & $\sim$1   & ------          \\ \hline
		5th method & 0.03516 & 0.7       & Tesla K80        \\ \hline
	\end{tabular}
	\caption{The quantitative results of official test set for NTIRE2020 ``Clean'' track.}
	\label{table4}
	\vskip -0.14in
\end{table}

\begin{table}[]
	\vskip 0.05in
	\centering
	\begin{tabular}{|l|l|l|l|}
		\hline
		Method    & MRAE   & Runtime/s & Compute Platform \\ \hline
		1st method     & 0.06201 & 3.748     & NVIDIA Titan XP    \\ \hline
		2nd method   & 0.06213 & 16        & NVIDIA 1080Ti    \\ \hline
		\textbf{AWAN+}    & \textbf{0.06217} & 0.56     &   NVIDIA 2080Ti\\ \hline
		4th method   & 0.06515 & $\sim$30   & NVIDIA Titan XP          \\ \hline
		5th method & 0.06733 & ------       & NVIDIA 2080Ti        \\ \hline
	\end{tabular}
	\caption{The quantitative results of official test set for NTIRE2020 ``Real World'' track.}
	\label{table5}
	\vskip -0.14in
\end{table}

\textbf{Adaptive weighted channel attention (AWCA).} Based on the baseline network, we conduct another experiment to inspect the effect of AWCA module. Results of $E_c$ and $E_h$ brings 5.0\% and 2.2\% decrease in MRAE metrics over the baseline results for the NTIRE2020 ``Clean" and ``Real World" tracks. The main reason lies in that AWCA module adaptively integrates channel-wise interdependencies for more powerful feature correlation learning. Afterwards, we combine PSNL and AWCA modules to further strengthen discriminant learning of our network. Experimental results of $E_d$ and $E_i$ demonstrate that the more superior performance can be achieved with such two modules.

\textbf{Camera spectral sensitivity (CSS) prior.} Experiments of $E_a$ to $E_d$ and $E_f$ to $E_i$ are all optimized by stochastic gradient descent algorithm with individual MRAE loss term $l_h$ in Section \ref{Camera Spectral Sensitivity (CSS) Prior}. Since the CSS function is known in the ``Clean" track and unknown in the ``Real World" track, we can only introduce CSS prior into AWAN network for ``Clean" track. $E_e$ means that we utilize a linear combination of MRAE loss term $l_h$ and CSS constraint $l_r$ as the final loss function and indicates that the incorporation of CSS prior is useful to improve the accuracy of spectral reconstruction.

\subsection{Results}
\label{Results}
To test the superiority of our proposed network, we compare our algorithm with six state-of-the-art methods, including Arad\cite{arad2016sparse}, Galliani\cite{galliani2017learned}, Yan\cite{yan2018accurate}, Stiebel\cite{stiebel2018reconstructing}, HSCNN-R\cite{shi2018hscnn+}, and HSCNN-D\cite{shi2018hscnn+}. The numerical results of validation set for NTIRE2020 and  NTIRE2018 ``Clean'' and ``Real World'' tracks are listed in Table \ref{table2} and Table \ref{table3}. As in \cite{shi2018hscnn+}, we also adopt multi-model ensemble method denoted as AWCA+. For the NTIRE2020 ``Clean'' track, additional three models are trained, including one model with 8 DRABs and $200$ output channels and two models with 20 DRABs and $128$ output channels. For the NTIRE2020 ``Real World'' track, we firstly adopt self-ensemble method for single AWAN network that the RGB input is flipped up/down to acquire a mirrored output and then the mirrored output and the original
output are averaged into the target result. Additional two models with 8 DRABs and $200$ output channels and one model with 10 DRABs and $180$ output channels are trained for multi-model ensemble. As for NTIRE2018 datasets, we perform the similar self-ensemble method as NTIRE2020 ``Real World'' track. Also, additional two models with 8 DRABs and $200$ output channels are on the ``Clean'' track and additional two models with 10 DRABs and $200$ output channels are for the ``Real World'' track. From Table \ref{table2} and Table \ref{table3}, we can observe that our single model outperforms other compared approaches and our method further improves the performance of SR with model-ensemble strategy.
Finally, our entries obtain the 1st ranking on the official test set of ``Clean" track and the 3rd place only 1.59106e-4 more than the 1st on the ``Real World" track in the NTIRE 2020 Spectral Reconstruction Challenge (see in Table \ref{table4} and Table \ref{table5}). It should be noted that we only list top 5 methods. 
\begin{figure}[htbp]
	\vskip 0.05in
	\centering
	\subfloat[]{\includegraphics[width=4.0cm,height=4.0cm]{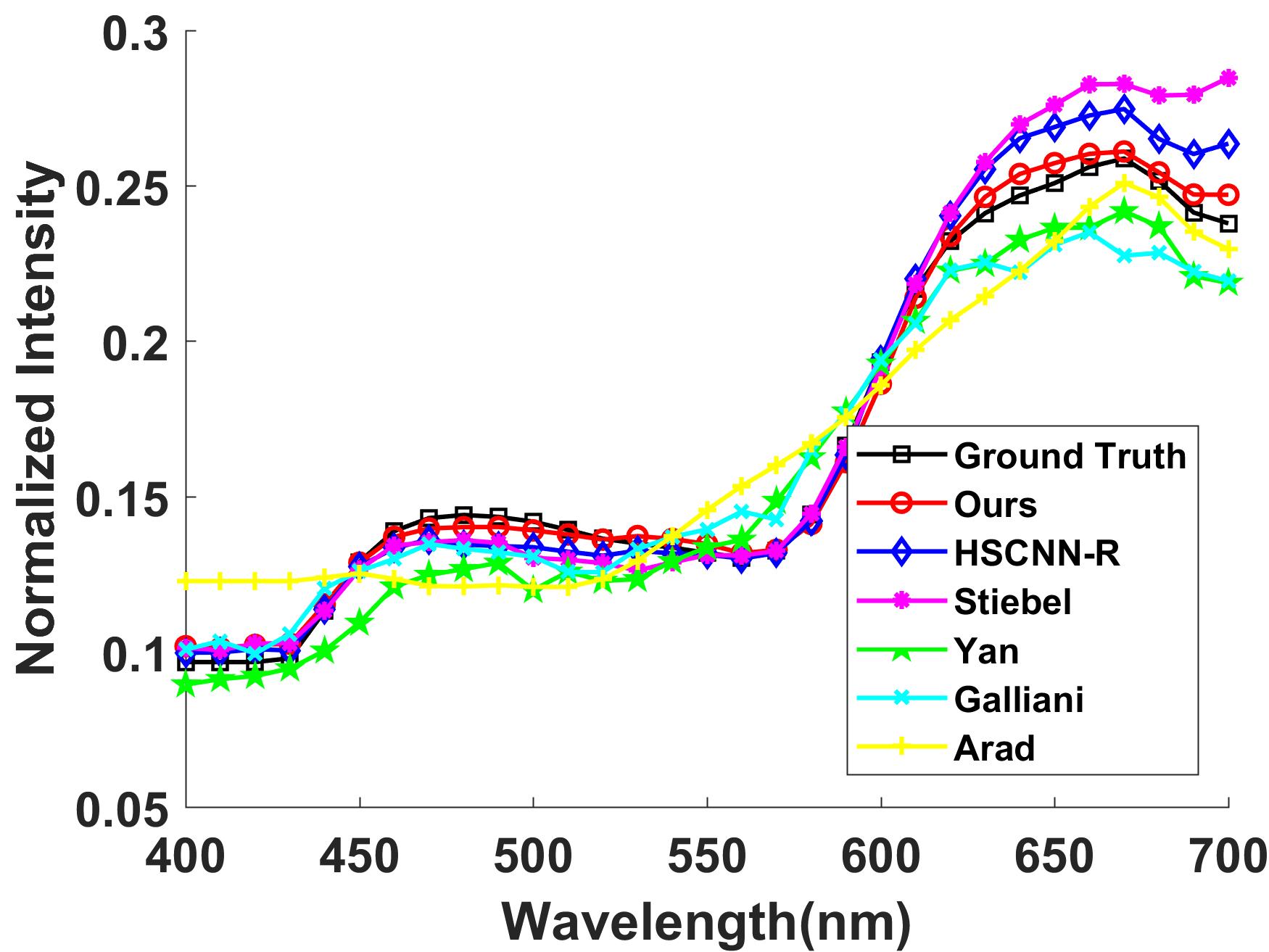}}\
	\subfloat[]{\includegraphics[width=4.0cm,height=4.0cm]{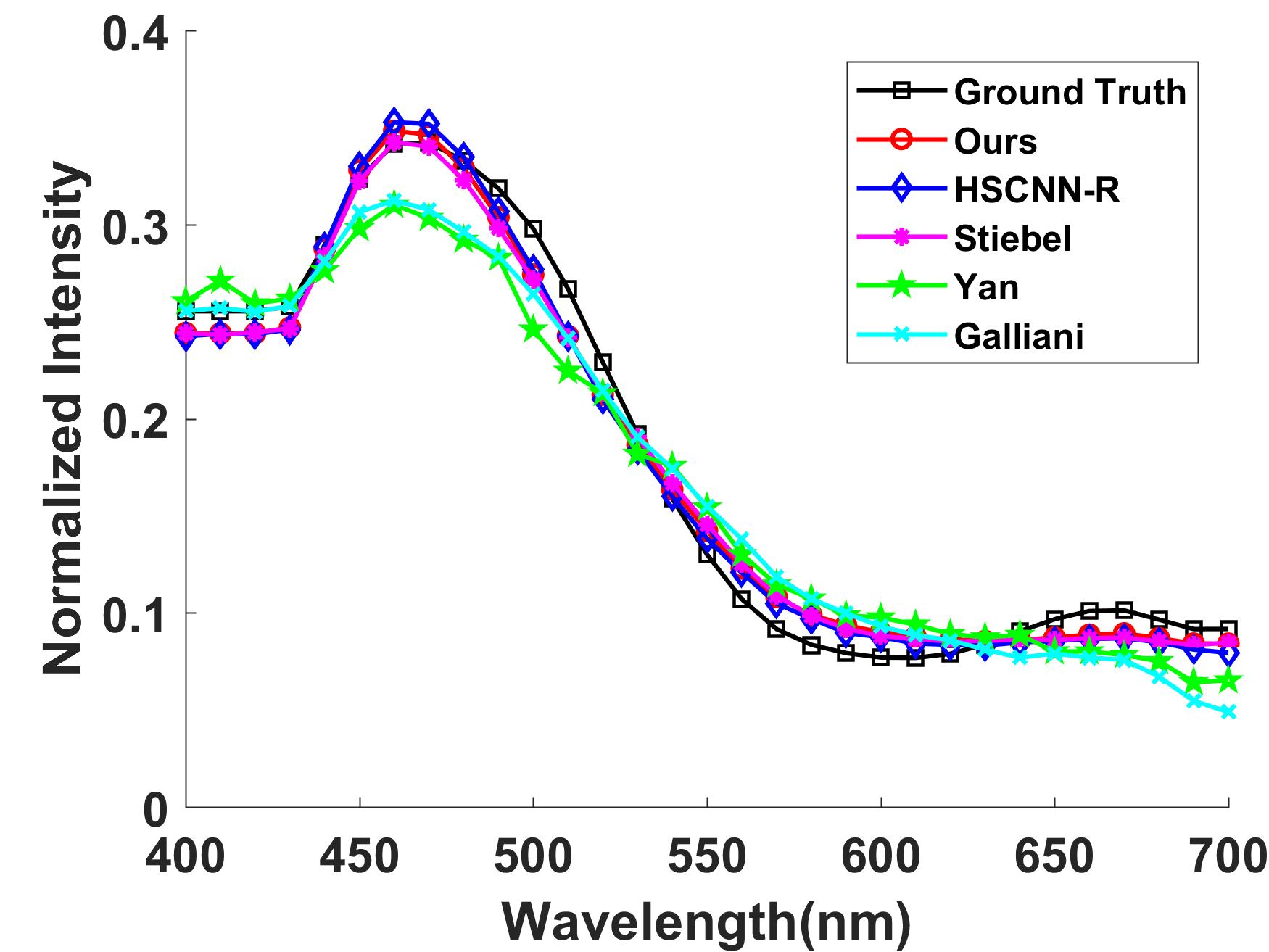}}\\
	\subfloat[]{\includegraphics[width=4.0cm,height=4.0cm]{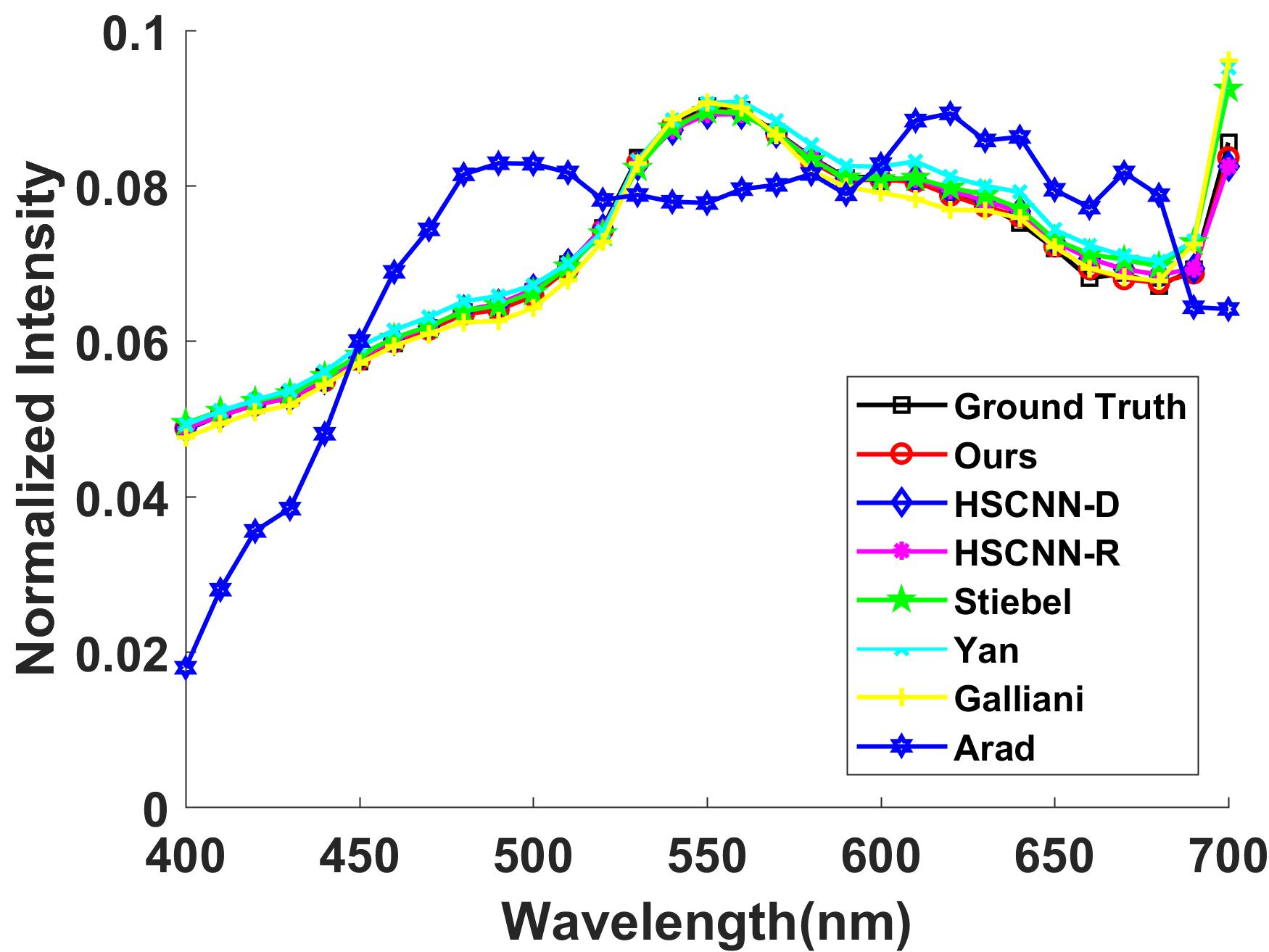}}\
	\subfloat[]{\includegraphics[width=4.0cm,height=4.0cm]{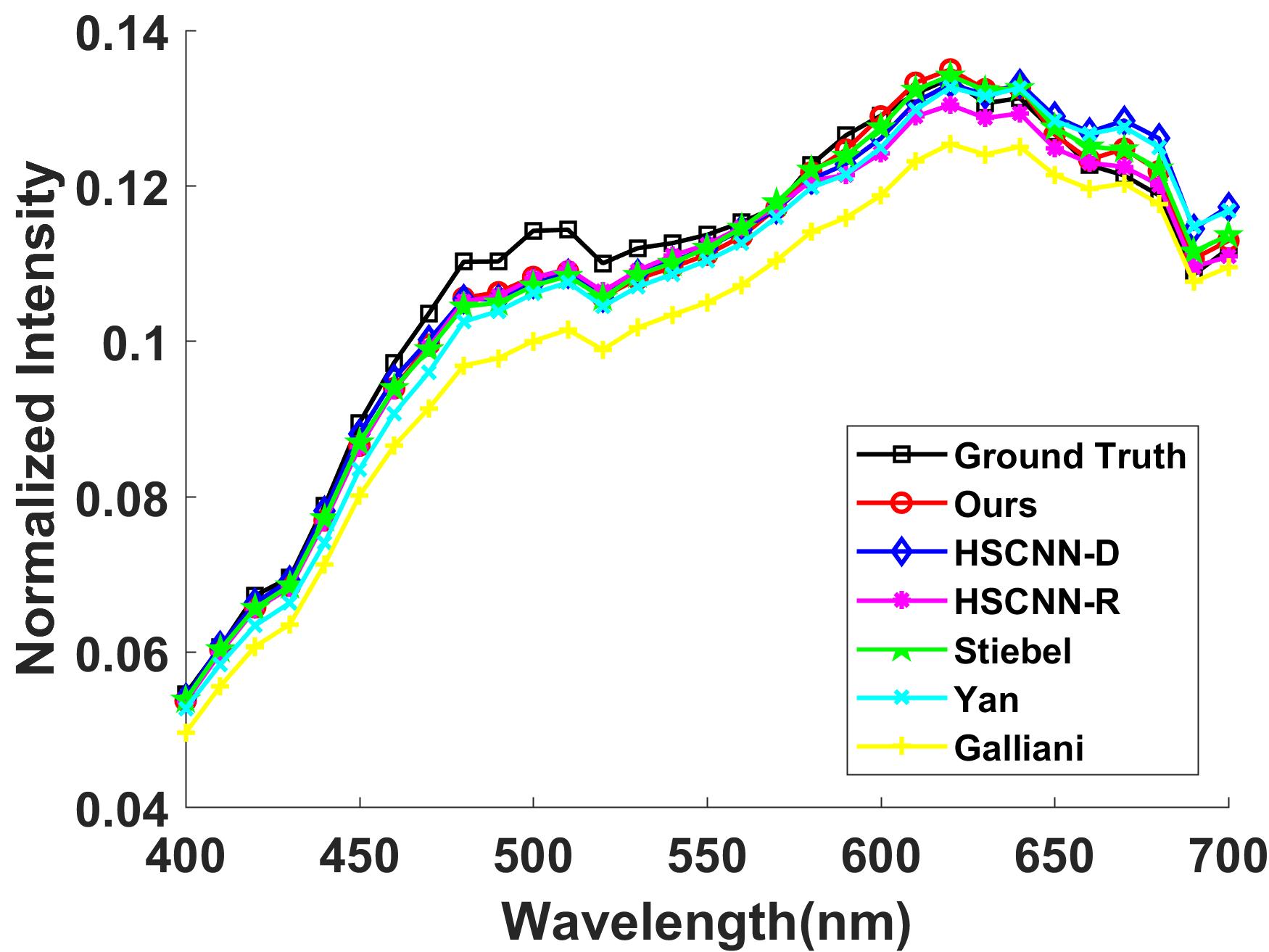}}\\
	\caption{Spectral response curves of selected several spatial points from the reconstructed HSI. (a) and (b) are for the NTIRE2018 ``Clean'' and ``Real World'' tracks respectively. (c) and (d) are for the NTIRE2020 ``Clean'' and ``Real World'' track respectively.}
	\label{figure8}
	\vskip -0.14in
\end{figure}

${\textbf{Visual results.}}$ To evaluate the perceptual quality of SR results, we display some visual reconstructed HSIs and the corresponding error maps of different methods in Figure \ref{figure5}, Figure \ref{figure7} and Figure \ref{figure6}. From these figures, we can see that our approach yields better recovery results and higher reconstruction fidelity than other methods. In addition, we also plot the spectral response curves in Figure \ref{figure8}. Obviously, the results of our proposed method are more accurate, which are closer to the ground truth HSIs. 

\section{Conclusion}
\label{Conclusion}
In this paper, we propose a deep adaptive weighted attention network (AWAN) for SR. Specifically, a patch-level second-order non-local (PSNL) module is presented to capture distant region correlations via second-order non-local operations. Besides, a trainable adaptive weighted channel attention (AWCA) module is proposed to adaptively recalibrate channel-wise feature responses by exploiting adaptive weighted feature statistics. To further improve the accuracy of SR, we introduce camera spectral sensitivity (CSS) prior and incorporate the discrepancies of the RGB images and HSIs as a finer constraint. Experimental results on challenging benchmarks demonstrate the superiority of our AWAN network in terms of numerical and visual results.

{\small
\bibliographystyle{ieee_fullname}
\bibliography{egbib}
}

\end{document}